\documentclass[12pt,a4paper]{article}
\usepackage[margin=2.5cm]{geometry}
\usepackage{amsmath,amssymb}
\usepackage{graphicx}
\usepackage{algorithm,algorithmic}
\usepackage{amsthm}
\usepackage{bm}
\usepackage[hidelinks]{hyperref}
\usepackage{authblk}

\newtheorem{theorem}{Theorem}[section]
\newtheorem{lemma}[theorem]{Lemma}

\title{Sequential Exchange Monte Carlo: A Sampling Method for Bayesian Data Analysis without Parameter Tuning}

\author[1]{Tomohiro Nabika}
\author[2]{Kenji Nagata}
\author[1]{Shun Katakami}
\author[3]{Masaichiro Mizumaki}
\author[1]{Masato Okada\thanks{okada@edu.k.u-tokyo.ac.jp}}
\affil[1]{Graduate School of Frontier Sciences, The University of Tokyo, Kashiwa, Chiba 277-8561, Japan}
\affil[2]{Research and Services Division of Materials Data and Integrated System, National Institute for Materials Science, Tsukuba, Ibaraki 305-0047, Japan}
\affil[3]{Japan Synchrotron Radiation Research Institute (JASRI), Sayo, Hyogo 679-5198, Japan}
\date{}

\begin{document}
\maketitle

\begin{abstract}
Bayesian data analysis is widely used across many disciplines, and representative examples in materials science include spectral analysis and sparse modeling. 
In such applications, the underlying models often become complex and yield multimodal posterior distributions, making efficient sampling from multimodal distributions essential. 
Replica exchange Monte Carlo has been commonly employed for this purpose; however, its performance strongly depends on difficult parameter tuning, such as the design of the inverse temperature. 
In this study, we comparatively investigate sampling algorithms that require fewer tuning parameters for Bayesian data analysis in materials science. 
Specifically, we compare three approaches: non-reversible parallel tempering (NRPT), sequential Monte Carlo samplers (SMCS), and a newly proposed method, sequential exchange Monte Carlo (SEMC). 
Our results indicate that NRPT can require computational time for parameter tuning, while SMCS requires careful adjustment of the number of MCMC steps at each temperature level. 
In contrast, SEMC achieves robust convergence across a range of problem settings without additional tuning, demonstrating its practicality for Bayesian inference.
\end{abstract}

\section{Introduction}
Bayesian inference is widely used for data analysis across various fields.
In materials science, a representative example is spectral data analysis \cite{Nagata2012}. 
By analyzing spectral data within the Bayesian framework, it becomes possible to select the number of peaks and to estimate peak parameters together with uncertainty quantification. 
Another representative application is sparse modeling \cite{nagata2015exhaustive}. 
Previous studies have reported that applying Bayesian sparse modeling to battery datasets enables the prediction of coordination energies of alkali-metal ions \cite{obinata2022data}. \par

It is well known that the posterior distribution in Bayesian inference can be multimodal when the model is complex.
In such cases, the random-walk Metropolis \cite{metropolis1953equation} algorithm, one of the most commonly used Markov chain Monte Carlo (MCMC) methods, often becomes inefficient, because the Markov chain can remain trapped near a single mode and rarely transitions between well-separated modes. 
To improve sampling from multimodal posteriors, replica exchange Monte Carlo (REMC), also known as parallel tempering, has been widely employed \cite{geyer1991markov, Hukushima1996}. REMC introduces a ladder of tempered distributions that interpolate between the prior and the posterior. 
By sampling each tempered distribution with the Metropolis algorithm and proposing exchanges between neighboring temperatures, REMC enhances global exploration and facilitates transitions across local minima. 
Despite its effectiveness, REMC typically requires careful tuning in practice, including the design of the inverse temperature and the choice of step sizes at each temperature level. 
Such manual parameter tuning hinders the use of the REMC method by non-experts.\par

In this study, we aim to comparatively investigate methods that enable Bayesian inference for complex models with little parameter tuning. 
The first method is non-reversible parallel tempering (NRPT) \cite{syed2022non}. 
NRPT extends REMC by introducing a deterministic exchange scheme and an automatic procedure to adjust the number of temperature levels and their values. 
When combined with an automatic step-size adaptation scheme, NRPT can largely eliminate the need for tuning parameters. 
However, these adaptations are typically performed during a burn-in phase, and depending on the problem setting, a sufficiently long burn-in period may be required to obtain stable tuning and reliable sampling. \par

The second method we consider is the sequential Monte Carlo sampler (SMCS) \cite{neal2001annealed, chopin2002sequential, del2006sequential}.
The SMCS method enables high-precision sampling from multimodal distributions by simultaneously handling a large number of samples.
Initially, many samples are drawn from the prior distribution, which is assigned the highest temperature distribution.
The algorithm then iteratively proposes the next temperature, resamples for the proposed distribution, and updates the samples using MCMC.
Depending on the problem, SMCS may require a large number of MCMC steps per sample.
Increasing the number of MCMC steps under a fixed computational budget generally reduces the number of parallel samples, which can degrade posterior estimation accuracy.
Therefore, it is crucial to tune the number of MCMC steps per sample to maintain sampling diversity while ensuring computational efficiency. \par

The third method is sequential exchange Monte Carlo (SEMC), which we newly propose as a hybrid of REMC and SMCS.
The algorithm begins by generating numerous samples at the highest temperature, similar to SMCS. 
It then iteratively proposes the next temperature, performs exchanges between distributions, and updates the parameters using MCMC. 
This algorithm does not require the long burn-in period, as in the SMCS method. 
In contrast to SMCS, where multiple MCMC chains often start from identical samples after the resampling step, SEMC ensures that MCMC updates are initialized from distinct samples, thereby enhancing the diversity of the samples. 
This allows SEMC to achieve high-precision sampling from multimodal distributions without the need for tuning the number of MCMC steps even when the number of intermediate distributions is small.

This study evaluated the performance of the three methods in two contexts: sampling from artificial multimodal distributions and performing Bayesian inference in materials science applications.
In the artificial multimodal sampling task, SEMC exhibited superior convergence compared with REMC, particularly when the number of intermediate distributions was small, and achieved higher accuracy than SMCS without requiring tuning of the MCMC steps.
Similar advantages were observed in the materials science application, demonstrating the practical effectiveness of SEMC in real-world inference problems.

The remainder of this paper is organized as follows.
Section 2 reviews the background of Bayesian data analysis and the sampling methods REMC and SMCS.
Section 3 introduces the proposed SEMC method.
Section 4 compares the performance of SEMC, NRPT, and SMCS through numerical experiments, and Section 5 discusses the results.
Finally, Section 6 concludes the paper and outlines directions for future work.

\section{Background}
This section describes data analysis by Bayesian inference with REMC and SMCS.
\subsection{Bayesian Inference}
Let $M$ be a model with parameter $\theta$ and data $D = \{X_i, y_i\}_{i=1}^N$.
Suppose that the likelihood function, $p(D|\theta, M)$, is given.
The error function, $E(\theta, M)$, representing the discrepancy between the data $D$ and the output of parameter $\theta$, is defined as follows:
\begin{align}
  \label{eq:likelihood}
  p(D|\theta, M) = \frac{\exp(-NE(\theta, M))}{C},
\end{align}
where $C$ is the normalization constant.
The goal of the data analysis is to estimate model $M$ and model parameter $\theta$ from $D$. \par

Assume that $\theta$ follows the prior distribution $p(\theta|M)$.
In Bayesian inference, the posterior probability, $p(\theta|D, M)$, is calculated as follows:
\begin{align}
  \label{eq:posterior}
  p(\theta|D, M) &= \frac{p(D|\theta, M)p(\theta|M)}{p(D|M)} \\
  &= \frac{\exp(-NE(\theta, M))p(\theta|M)}{\int \exp(-NE(\theta, M))p(\theta|M)d\theta}.
\end{align}
Let $\mathcal{M}$ be the set of candidate models, and let $p(M)$ be the prior probability of model $M \in \mathcal{M}$.
The posterior probability of model $M$ is given by
\begin{align}
  \label{eq:model_posterior}
  p(M|D) &= \frac{p(D|M)p(M)}{p(D)} \\
         &= \frac{p(M)}{\overline{Z}} \exp(-F(M)) \\
  F(M) &= -\log\int \exp(-NE(\theta, M))p(\theta|M)\textup{d}\theta + \log C, \\
  \overline{Z} &= \sum_{M' \in \mathcal{M}} p(M')\exp(-F(M')),
\end{align}
where $F(M)$ is the Bayesian free energy, which represents the goodness of model $M$ for $D$. \par
By computing the posterior probabilities $p(\theta|D, M)$ and $p(M|D)$,  $M$ and $\theta$ can be estimated.
Since analytical calculation of these posterior probabilities is generally difficult, MCMC methods are commonly used. 
The following sections describe two conventional MCMC algorithms: REMC and SMCS.

\subsection{Replica Exchange Monte Carlo}
The REMC method was originally developed for physical simulation \cite{geyer1991markov, Hukushima1996}
 and has since been applied to various Bayesian inference problems \cite{von2011bayesian, sambridge2014parallel}.
 In this section, we describe the REMC algorithm in the context of Bayesian inference. \par

\subsubsection{Algorithm}
In the analysis of complex models, $p(\theta|D, M)$ is often multimodal, which makes it prone to becoming trapped in local solutions and challenging to sample accurately.
REMC mitigates this issue by preparing multiple probability distributions at different temperatures.
Let $E(\theta, M)$ denote the error function; the distribution at an inverse temperature $\beta$ is defined as follows:
\begin{align}
  \label{eq:temperature}
  p_\beta(\theta) &\propto p(\theta|D,M)^{\beta}p(\theta|M)^{1-\beta} \\
  &\propto  \exp(-\beta N E(\theta, M))p(\theta|M).
\end{align}
Let $L$ be the number of distributions, and let $(\beta_l)_{l=1}^L$ be the set of inverse temperatures. (Here, $0 = \beta_1 < \cdots < \beta_L = 1$.)
Under the REMC method, the following two parameter-update procedures are repeatedly performed to sample from $(p_{\beta_l})_{l = 1}^L$:
\begin{enumerate}
  \item Update parameter $\theta_l$ according to $p_{\beta_l}$ using the MCMC algorithm.
  \item Exchange parameters $\theta_l$ and $\theta_{l+1}$ between $p_{\beta_{l}}$ and $p_{\beta_{l+1}}$ with a probability of $u$.
  \begin{align}
    u = \min\left(1, \frac{p_{\beta_{l}}(\theta_{l+1})p_{\beta_{l+1}}(\theta_{l})}{p_{\beta_{l}}(\theta_{l})p_{\beta_{l+1}}(\theta_{l+1})}\right).
  \end{align}
\end{enumerate}
The specific algorithm is provided in Appendix\ref{sec:appendix_algorithm}. \par

By sampling from the distribution with inverse temperature $\beta_L = 1$, samples following the posterior distribution $p(\theta|D, M)$ are obtained.
The posterior distribution can then be approximated using a histogram or kernel density estimation \cite{PRML}. \par
Calculation of the Bayesian free energy, $F(M)$, is required to evaluate the model posterior distribution.
This calculation can be efficiently performed using all samples from all distributions \cite{neal1993probabilistic}.
Let $z(\beta)$ be defined as follows:
\begin{align}
  z(\beta) = \int \exp(-\beta NE(\theta, M))p(\theta|M)d\theta.
\end{align}
Thus, the Bayesian free energy, $F(M)$, was calculated as follows:
\begin{align}
  F(M) &= -\log z(1) + \log C, \\
  z(1)  &= \frac{z(\beta_L)}{z(\beta_{L-1})} \times \cdots \times \frac{z(\beta_2)}{z(\beta_1)} = \prod_{l=1}^{L-1} \frac{z(\beta_{l+1})}{z(\beta_l)} \\
       &= \prod_{l=1}^{L-1} \frac{\int \exp(-\beta_{l+1} NE(\theta, M))p(\theta|M)}{\int \exp(-\beta_l NE(\theta, M))p(\theta|M)} \\
       &= \prod_{l=1}^{L-1} \langle\exp(-N(\beta_{l+1} - \beta_{l})E(\theta,M))\rangle_{p_{\beta_l}(\theta)},
\end{align}
where $\langle \cdot \rangle_{p_{\beta_l}(\theta)}$ denotes the expectation value with probability $p_{\beta_l}(\theta)$.
\subsubsection{Exchange Scheme}
Various strategies have been proposed for selecting pairs of intermediate distributions for exchange moves.
In the stochastic even-odd (SEO) scheme, adjacent distribution pairs 
$\{(l-1,l)\}_{l=2}^L$ are divided into two subsets:
\begin{align}
  E &= \{(l-1,l) \mid l \textup{ is even}\}, \\
  O &= \{(l-1,l) \mid l \textup{ is odd}\}.
\end{align}
At each exchange step, either the $E$ or $O$ subset is randomly selected with equal probability, and exchanges are attempted for all pairs within the chosen subset.
In NRPT, which is an extension of REMC, the deterministic even-odd (DEO) scheme alternates deterministically between the $E$ and $O$ subsets at each step \cite{syed2022non}.
Although DEO satisfies only the skew-detailed balance condition, it achieves superior performance, as indicated by the round-trip rate:
$O(L^2)$ for SEO and $O(L)$ for DEO, where $L$ is the number of intermediate distributions. \par

In REMC, the number of intermediate distributions $L$, the inverse temperatures $(\beta_l)_{l=2}^L$, and their step sizes are crucial parameters that strongly affect algorithm performance; various tuning strategies have been proposed in the literature \cite{kone2005selection, miasojedow2013adaptive}.
For the DEO scheme, it is known that setting the exchange probability to 1/2 leads to optimal round-trip rates.
In NRPT, automatic tuning for the inverse temperatures and the number of intermediate distributions is performed during a burn-in period so that the exchange probability approaches 1/2 \cite{syed2022non}.
 \par


\subsubsection{MCMC Kernel}
Within the REMC framework, each sample is updated using an MCMC kernel after the exchange step.
The choice of MCMC kernel significantly affects the overall convergence of REMC.
Various kernels, including the Metropolis algorithm, Gibbs sampling, Hamiltonian Monte Carlo (HMC) \cite{neal2011mcmc}, and Langevin dynamics \cite{roberts2002langevin}, have been developed to enhance sampling efficiency across different problem settings.
In this study, to ensure a fair and interpretable comparison with SMCS and SEMC, we employ the most basic and widely used MCMC kernel: the random-walk Metropolis algorithm. \par

In the random-walk Metropolis algorithm, step-size tuning is essential for effective exploration of the parameter space.
If the step size is too small, proposed samples remain close to the current state, resulting in slow mixing.
Conversely, if the step size is too large, proposals are frequently rejected due to low acceptance probabilities.
Therefore, appropriate step sizes must be selected for each parameter. \par

When all parameters are updated simultaneously in a single Metropolis move, assigning optimal step sizes individually becomes challenging.
To address this, we adopt a sequential update scheme in which parameters are updated one at a time.
This allows the acceptance rate of each parameter to be measured independently, enabling adaptive adjustment of step sizes based on each parameter's acceptance behavior.
We employ the Robbins-Monro algorithm \cite{robbins1951stochastic, garthwaite2016adaptive}
and the dual averaging method \cite{nesterov2009primal, hoffman2014no} to adaptively tune step sizes during sampling.
Further details of the adaptation procedure are provided in Appendix\ref{sec:appendix_tuning}. \par

\subsection{Sequential Monte Carlo Samplers}
\subsubsection{Algorithm}
The SMCS method \cite{neal2001annealed, chopin2002sequential, del2006sequential} enables sampling from multimodal distributions by simultaneously managing multiple samples.
Similar approaches, such as transitional Markov chain Monte Carlo (TMCMC) \cite{ching2007transitional} and population annealing \cite{hukushima2003population} have been developed and applied across various fields \cite{lye2021sampling,weigel2021understanding}. 
In this section, we describe the SMCS algorithm with MCMC updates for Bayesian inference. \par
In the SMCS method, samples following the probability distributions at inverse temperatures, $(\beta_l)_{l=1}^L$, are obtained by performing the following parameter updates:
\begin{enumerate}
  \item Let $T$ be the number of samples and $l = 1, \beta_1 = 0$. Generate samples $(\theta_1^i)_{i=1}^T$ according to the prior distribution.
  \item Repeat the following until $\beta_l = 1$.
  \begin{enumerate}
    \item Update $l = l + 1$.
    \item Resample all the samples according to the weight $(W_l^i)_{i = 1}^T$.
    \item Update $\theta_l^i$ using the MCMC algorithm with step size $\epsilon_l$. Repeat this step $n$ times.
    \item Set the next inverse temperature $\beta_{l + 1}$.
  \end{enumerate}
\end{enumerate}

Here, we defined $W_l^i = \exp(-(\beta_{l} - \beta_{l-1})NE(\theta_{l-1}^i))$. \par

\subsubsection{Waste-free SMC Algorithm}
In the SMCS method, the number of MCMC steps per sample must be sufficiently large to maintain sample diversity.
However, increasing the number of MCMC steps also increases computational costs.
To address this, the waste-free SMC method \cite{dau2022waste} was introduced, which efficiently utilizes the intermediate MCMC steps to improve sampling efficiency.\par

Waste-free SMC reduces inefficiency by resampling only $S$ samples and updating each sample $n$ times using the MCMC method.
This process generates $S$ chains, each of length $n$, which are combined to form a new particle sample of size $T = S \times n$. 
The detailed algorithm is provided in Appendix\ref{sec:appendix_algorithm}. 
This approach allows an increase in the number of steps without additional computational cost, thereby improving efficiency.
However, if $S$ is too small, the uncertainty within the resampling scheme can become significant.
Therefore, appropriate selection of $S$ is essential in practical applications.
\par

\subsubsection{Resampling Scheme}
In SMCS, various resampling schemes have been proposed to mitigate particle degeneracy \cite{chopin2022resampling}.
Multinomial resampling is the simplest and most widely used approach.  
In this scheme, for each index $i$, an ancestor index $j$ is selected with probability proportional to its normalized weight, as follows:
\begin{equation}
\theta_l^i = \theta_{l-1}^j \quad \text{with probability} \quad \frac{W_l^j}{\sum_j W_l^{j}}.
\end{equation}
Although easy to implement, multinomial resampling can suffer from high variance. \par

To address this, more advanced strategies, such as Srinivasan sampling process (SSP) resampling, have been developed \cite{gerber2019negative}.  
In SSP, the number of offspring for each particle is either $\left\lfloor T \frac{W_l^{j}}{\sum_j W_l^{j}} \right\rfloor$ or $\left\lfloor T \frac{W_l^{j}}{\sum_j W_l^{j}} \right  \rfloor + 1$, which reduces the variance and improves the accuracy of the resampling process. \par

In the waste-free SMC, which we employ in this study, the number of resampled particles is smaller than the number of original particles.  
Furthermore, multinomial resampling is the only scheme for which the theoretical validity of waste-free SMC has been rigorously established.  
Therefore, we adopt multinomial resampling in this work.

A common approach for determining the inverse temperatures is to set them so that the effective sample size (ESS), computed from the importance weights at $l-1$, equals $50\%$ of the total sample size.  
In this study, however, we adopt the same temperature selection procedure described in Section~3 to simplify computational comparisons.  
\subsubsection{MCMC Kernel}
As in REMC, various MCMC kernels can also be used within the SMCS framework to update the samples.  
Common choices include the Metropolis algorithm, Gibbs sampling, HMC \cite{burda2023hamiltonian}, and Langevin dynamics \cite{arampatzis2018langevin}.  
In this study, to enable a clear comparison with REMC and SEMC, we employ the random-walk Metropolis algorithm. \par

In the random-walk Metropolis algorithm, step-size tuning is critical for effective exploration of the parameter space.
A common approach is to set the step size based on the empirical variance of each parameter across the particle population \cite{ching2007transitional, safta2020transitional}.
This method generally performs well when the target distribution is unimodal or moderately dispersed.  
However, for multimodal distributions, the parameter variance across samples can become extremely large due to mode separation.  
To address this issue, we apply the same adaptive tuning scheme used in SEMC, as described in Section 3.  

\section{Sequential Exchange Monte Carlo}
This section describes the proposed SEMC method, which combines the REMC and SMCS methods.

\subsection{Algorithm}
By performing the following parameter updates, samples were obtained from the probability distributions with inverse temperatures $(\beta_l)_{l=1}^L$. \par
\begin{enumerate}
  \item Let $l = 1$ and $\beta_1 = 0$. Generate samples $(\theta_l^i)_{i=1}^T$ according to the prior distribution.
  \item Repeat the following until $\beta_l = 1$.
  \begin{enumerate}
    \item Update $l = l + 1$. Set the initial sample, $\theta^1_l = \theta_{l-1}^j$, with probability $\frac{W_l^j}{\sum_j W_l^j}$.
    \item Repeat the following update for sampling.
    \begin{enumerate}
      \item Update $\theta_l^i$ according to $p_{\beta_l}(\theta)$ using the MCMC method.
      \item Select sample $\theta_{l-1}^j$ randomly. Exchange $\theta_{l-1}^j$ and $\theta_l^i$ with probability $u$.
      \begin{align}
          u = \min\left(1, \dfrac{p_{\beta_{l}}(\theta_{l-1}^j)p_{\beta_{l-1}}(\theta_l^i)}{p_{\beta_{l}}(\theta_l^i)p_{\beta_{l-1}}(\theta_{l-1}^j)}\right).
      \end{align}
    \end{enumerate}
  \end{enumerate}
\end{enumerate}
The specific algorithm is provided in Appendix\ref{sec:appendix_algorithm}.
The computation of the SEMC method can be parallelized. For
the parallelization, multiple initial samples are selected, and each sample is updated
independently and in parallel using the exchange and MCMC steps. \par

\subsection{Setting the Inverse Temperatures}
We considered tuning $\beta_l$ from the samples, $\{\theta_{l-1}^i\}_{i=1}^T$.
The exchange rate, $J$, between $p_{\beta_{l-1}}$ and $p_{\beta_l}$, was defined as
\begin{align}
    J &= \int \int u p_{\beta_{l-1}}(\theta_1)p_{\beta_l}(\theta_2)\textup{d}\theta_1 \textup{d}\theta_2. \\
    u &= \min\left(1, r\right). \\
    r &= \dfrac{p_{\beta_l}(\theta_1)p_{\beta_{l-1}}(\theta_2)}{p_{\beta_l}(\theta_2)p_{\beta_{l-1}}(\theta_1)}.
\end{align}
$J$ can be transformed as follows:
\begin{align}
    J &= \int\int_{E(\theta_2) < E(\theta_1)} r p_{\beta_l}(\theta_2)p_{\beta_{l-1}}(\theta_1)\textup{d}\theta_1 \textup{d}\theta_2  + \int\int_{E(\theta_2) \geq E(\theta_1)} p_{\beta_l}(\theta_2)p_{\beta_{l-1}}(\theta_1)\textup{d}\theta_1 \textup{d}\theta_2 \\ 
    &= 2\int\int_{E(\theta_2) \geq E(\theta_1)} p_{\beta_l}(\theta_2)p_{\beta_{l-1}}(\theta_1)\textup{d}\theta_1 \textup{d}\theta_2 \\
    &= 2\frac{C_1}{C_2}\int\int_{E(\theta_2) \geq E(\theta_1)} \exp(-(\beta_l - \beta_{l-1})NE(\theta_1))p_{\beta_{l-1}}(\theta_2)p_{\beta_{l-1}}(\theta_1)\textup{d}\theta_1 \textup{d}\theta_2.
    \label{eq:exchange_rate}
\end{align}
$C_1$ and $C_2$ are the normalization constants of $p_{\beta_{l-1}}$ and $p_{\beta_l}$, respectively, and $\frac{C_2}{C_1}$ can be transformed as follows:
\begin{align}
    \frac{C_2}{C_1} &= \frac{\int \exp(-\beta_{l}NE(\theta))p(\theta)\textup{d}\theta}{\int \exp(-\beta_{l-1}NE(\theta))p(\theta)\textup{d}\theta}. \\
    &= \int \exp(-(\beta_{l} - \beta_{l-1})NE(\theta))p_{\beta_{l-1}}(\theta)\textup{d}\theta.
    \label{eq:ratio}
\end{align}
Equations (\ref{eq:exchange_rate}) and (\ref{eq:ratio}) show that $J$ can be approximated from the samples of inverse temperature $\beta_{l-1}$.
Thus, the inverse temperature parameter, $\beta_l$, can be set such that $J$ is constant. \par

\subsection{MCMC Kernel}
As in the REMC and SMCS methods, we use the Metropolis algorithm as the MCMC kernel. 
In this study, we propose a step-size tuning method for the Metropolis algorithm, based on the theoretical result\cite{nagata2024algebraic}. 
According to Theorem 1, the average acceptance rate $U_{l,i}$ for the parameter $\theta_{l,i}$ has the following asymptotic form when $N\beta_l$ and the step size $\sigma_{l,i}$ are sufficiently large:

\begin{align}
U_{l,i} \sim \frac{(\log N\beta_l)^{m_i - m}}{(N\beta_l)^{\lambda_i - \lambda}} \cdot \frac{c}{\sigma_{l,i}},
\label{eq:acceptance_rate}
\end{align}

where
\begin{itemize}
  \item $\lambda$ and $m$ are the pole and its order of the zeta function $\zeta(z) = \int E(\theta)^z p(\theta) \, d\theta$,
  \item $\lambda_i$ and $m_i$ are those of the zeta function $\zeta_i(z) = \int |\theta_i| E(\theta)^z p(\theta) \, d\theta$,
  \item $\sigma_{l,i}$ is the step size of the proposal distribution for $\theta_{l,i}$,
  \item $c$ is a constant depending on $E(\theta)$ and $p(\theta)$.
\end{itemize}

Based on Eq.~\eqref{eq:acceptance_rate}, we tune the step size $\sigma_{l,i}$ so that the average acceptance rate stays approximately constant at a target value $\alpha$.

Under the assumption that $m_i = m$, Eq.~\eqref{eq:acceptance_rate} shows the following relation:
\begin{align}
\sigma_{l,i} : \sigma_{l-1,i} : \sigma_{l-2,i} \sim 
\frac{1}{(\beta_l)^{\lambda_i - \lambda} U_{l,i}} : \frac{1}{(\beta_{l-1})^{\lambda_i - \lambda} U_{l-1,i}} : \frac{1}{(\beta_{l-2})^{\lambda_i - \lambda} U_{l-2,i}}.
\label{eq:step_scaling}
\end{align}

Therefore, the step size can be set explicitly using the previous step sizes and the average acceptance rates, as follows:

\begin{align}
\sigma_{l,i} &= \begin{cases}
\sigma_{l-1,i} \text{ if } l = 2, \\
\frac{A}{\alpha(\beta_l)^d} \text{ if } l \geq 3, \\
\end{cases} 
\\
d &= \frac{\log(\sigma_{l-2,i} U_{l-2,i}) - \log(\sigma_{l-1,i} U_{l-1,i})}{\log(\beta_{l-1}) - \log(\beta_{l-2})}, \\
A &= U_{l-1,i} \sigma_{l,i} (\beta_{l-1})^d.
\label{eq:step_setting}
\end{align}

\subsection{Theoretical Validity}

This section provides a theoretical justification for the proposed SEMC method in continuous state spaces, assuming the random-walk Metropolis algorithm is employed as the MCMC kernel. We analyze the asymptotic convergence of the empirical distributions generated by SEMC to the target distributions at each intermediate distribution.

Let the sample space be a subset of $\mathbb{R}^d$, and define a partition $\Delta^k = \{[n_1 2^{-k}, (n_1+1)2^{-k}) \times [n_2 2^{-k}, (n_2+1)2^{-k}) \times \ldots \times [n_d 2^{-k}, (n_d+1)2^{-k})\}_{n_1, n_2, \ldots, n_d \in \mathbb{Z}}$ of the space.
Two probability distributions $p$ and $q$ are defined over the partition $\Delta^k$, and we define the distance between them as:
\begin{align}
\|p - q\|_{\Delta^k} = \sup_{A \in \mathcal{B}(\Delta^k)} |p(A) - q(A)|,
\end{align}
where $\mathcal{B}(\Delta^k)$ denotes the Borel $\sigma$-algebra over partition $\Delta^k$.

Let $\{ \theta_l^i \}_{i=1}^N$ be the samples obtained at the $l$-th intermediate distribution in SEMC. The empirical distribution is defined as:
\begin{align}
p_l^{N} = \frac{1}{N} \sum_{i=1}^N \delta_{\theta_l^i},
\end{align}
where $\delta_{\theta_l^i}$ is the Dirac measure at $\theta_l^i$.
Assuming that $p_{\beta_l}$ is continuous, we obtain the result discussed next.

\begin{theorem}
    Let $\{ \theta_l^i \}_{i=1}^N$ be the samples obtained at the $l$-th intermediate distribution via SEMC. Then, the empirical distribution $p_l^{N}$ converges in probability to the target distribution $p_{\beta_l}$ on partition $\Delta^k$. 
    That is, for every $0 < \delta < 1$ and $\epsilon > 0$, $K \in \mathbb{N}$ exists such that for all $k \geq K$, $N_0 \in \mathbb{N}$ exists such that for all $N \geq N_0$,
    \begin{align}
    \Pr\left( \|p_l^{N} - p_{\beta_l}\|_{\Delta^k} < \epsilon \right) > \delta.
    \end{align}
\end{theorem}

\paragraph{Sketch of proof.}
Let $Q_l$ denote the Markov kernel of the $l$-th intermediate distribution in SEMC. Define $Q_l^{\text{hist}}$ as the discretized version of $Q_l$ over the partition $\Delta^k$. We also define an idealized kernel $\widetilde{Q}_l$ where samples from layer $l-1$ are drawn from the exact distribution $p_{\beta_{l-1}}$, and its discretized version $\widetilde{Q}_l^{\text{hist}}$.

Let the stationary distributions of these kernels $Q_l, Q_l^{\text{hist}}, \widetilde{Q}_l, \widetilde{Q}_l^{\text{hist}}$ be denoted by $\pi_l$, $\pi_l^{\text{hist}}$, $\widetilde{\pi}_l$, and $\widetilde{\pi}_l^{\text{hist}}$. From the detailed balance, we obtain $\widetilde{\pi}_l = p_{\beta_l}$, and by construction, $\|\widetilde{\pi}_l - \widetilde{\pi}_l^{\text{hist}}\|_{\Delta^k} = 0$.

Both $\|\pi_l - \pi_l^{\text{hist}}\|_{\Delta^k}$ and $\|\pi_l^{\text{hist}} - \widetilde{\pi}_l^{\text{hist}}\|_{\Delta^k}$ become arbitrarily small as $k$ and $N$ are increased.
Furthermore, the empirical distribution $p_l^{N,k}$ generated by $Q_l$ converges to $\pi_l$, and thus to $p_{\beta_l}$, in distance $\| \cdot \|_{\Delta^k}$.

A full formal proof is provided in Appendix\ref{sec:appendix_proof}.

\section{Numerical Experiments}
We evaluated the performance of SEMC, NRPT, and waste-free SMC through numerical experiments, comparing computational time based on the number of evaluations of the energy function $E(\theta)$.
Section 4.1 presents the results for sampling from a multimodal distribution, and Section 4.2 presents the results for Bayesian inference in materials science.

\subsection{Sampling from a Multimodal Distribution}
\subsubsection{Problem Setting}
\begin{figure}[h]
  \centering
  \includegraphics[width = 8cm]{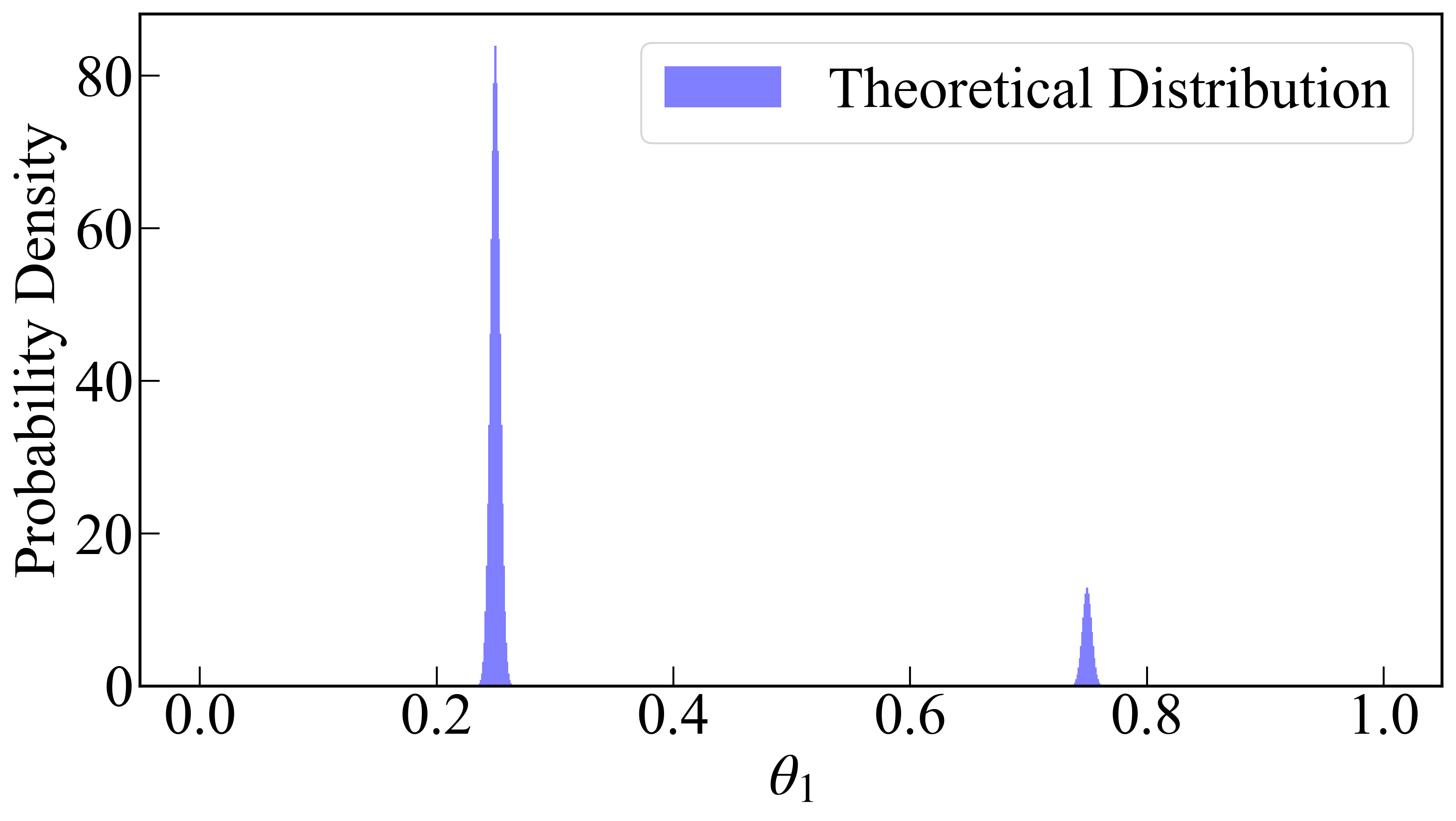}
  \caption{Theoretical histogram of $\theta_1$ with a bin width of 0.001.}
  \label{fig:multimodal}
\end{figure}

We defined the prior distributions as a uniform distribution on $[0,1]$ for $\theta_1$, and standard normal distributions for $\{\theta_2, \ldots, \theta_d\}$.
We defined the error function $E(\theta)$ as follows:
\begin{align}
  E_r(\theta ) = \begin{cases}
      30030(\theta_1 - 0.25)^2 +300 \left( \sum_{i=2}^{d} (\theta_i)^2 + 2r\sum_{i=2}^{d}\sum_{j=i+1}^{d} \theta_i \theta_j \right) & (\theta_1 < 0.5) \\
      30000(\theta_1 - 0.75)^2 +300 \left( \sum_{i=2}^{d} (\theta_i)^2 + 2r\sum_{i=2}^{d}\sum_{j=i+1}^{d} \theta_i \theta_j \right) + \frac{15}{8}. & (\theta_1 > 0.5)
  \end{cases}           
\end{align}
Figure \ref{fig:multimodal} shows the theoretical values of a histogram of $\theta_1$ with a bin width of 0.001. 
Error function $E(\theta)$ is continuous in this problem setting, ensuring the theoretical validity of SEMC. 
Furthermore, the parameter $r$ controls the correlation among the parameters ${\theta_2, \ldots, \theta_d}$, allowing us to examine how each method's performance varied with the correlation structure.
For each value of $r$, the free energy $F_r$ can be computed analytically, enabling direct comparison with the results obtained through sampling.
The derivation of $F_r$ is provided in Appendix\ref{sec:appendix_problem_setting}.
 \par
\subsubsection{Tuning Parameters}
\begin{figure}[h]
  \centering
  \includegraphics[width=16cm]{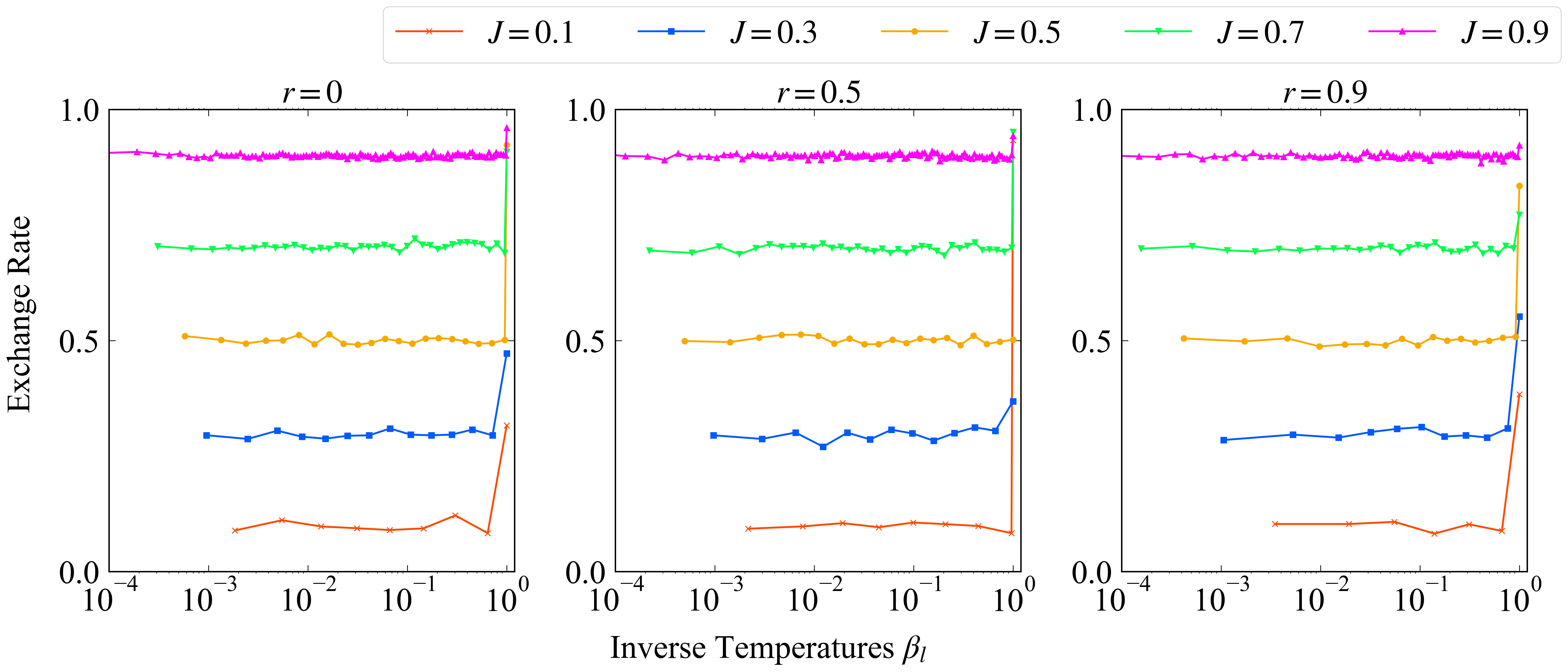}
  \caption{Inverse temperature $(\beta_l)_{l=2}^L$, and exchange rate between the $p_{\beta_l}, p_{\beta_{l-1}}$ of the SEMC method with $J = 0.1, 0.3, 0.5, 0.7, 0.9$.
  From left to right, the results are shown for correlation parameters $r = 0, 0.5, 0.9$.
  }
  \label{fig:exchange_multi_SEMC}
\end{figure}

This section presents the results of parameter tuning for SEMC.
We set the number of samples to 6000, $r = 0$, and the target acceptance rate to $\alpha = 0.5$.
The target exchange rate $J$ was varied across $0.1$, $0.3$, $0.5$, $0.7$, and $0.9$. Figure \ref{fig:exchange_multi_SEMC} shows the resulting actual exchange rates.
The results indicate that the exchange rates are accurately maintained across most temperatures, except between $\beta_L$ and $\beta_{L-1}$.
This deviation occurs because, when the computed inverse temperature exceeds 1, it is capped at 1.
\par

\begin{figure}[h]
  \centering
  \includegraphics[width=16cm]{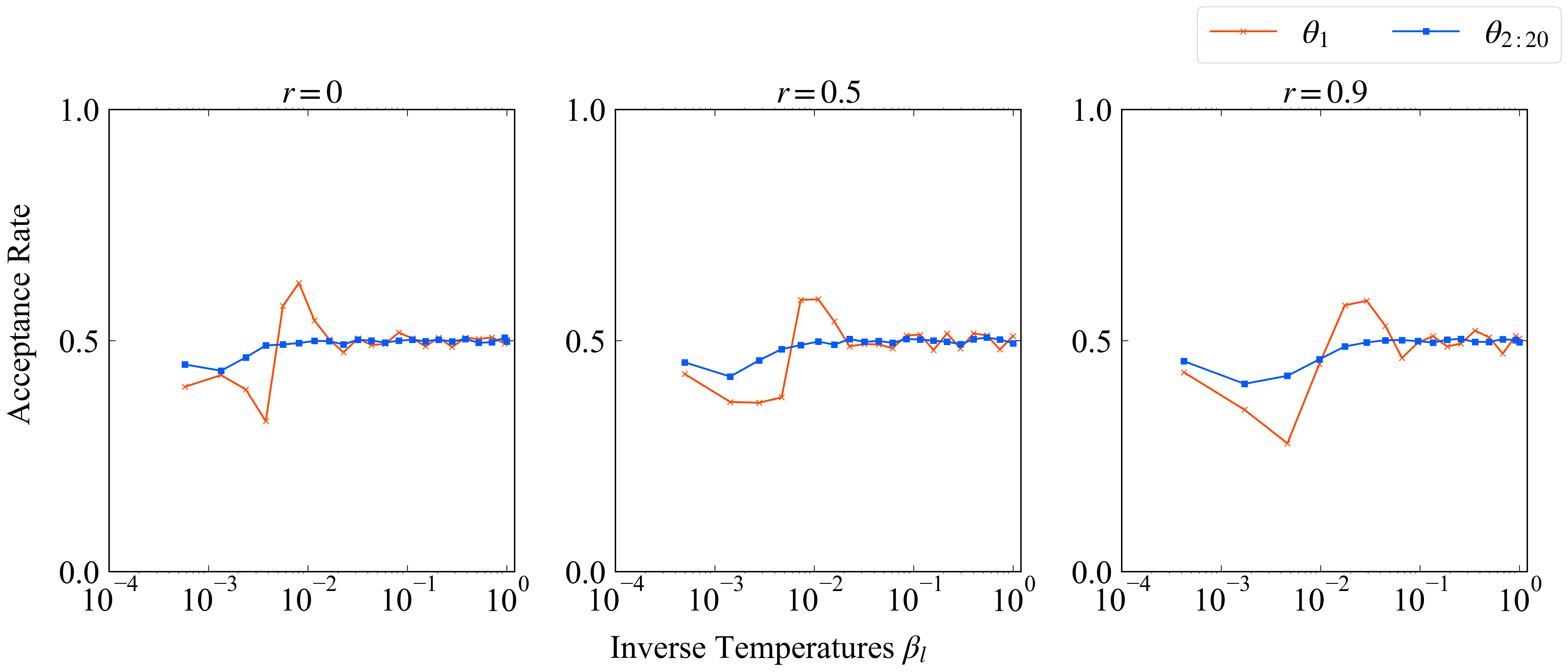}
  \caption{Average acceptance rates of the Metropolis algorithm at each inverse temperature for SEMC with $J = 0.5$.
  From left to right, results are shown for correlation parameters $r = 0, 0.5, 0.9$.
  The red line indicates the acceptance rate for $\theta_1$, and the blue line represents the average acceptance rate for $\theta_2, \ldots, \theta_d$.
  }
  \label{fig:acceptance_multi_SEMC}
\end{figure}

We also examined how well the Metropolis acceptance rates are controlled under the settings $\alpha = 0.5$ and $J = 0.5$, using 6000 samples.
Figure \ref{fig:acceptance_multi_SEMC} shows the average acceptance rates of the Metropolis algorithm at each inverse temperature.
The results indicate that approximate Equation (\ref{eq:step_setting}) accurately predicts the behavior of the acceptance rates, particularly in regions with larger $\beta_l$.

\subsubsection{Comparison with NRPT}
\begin{figure}[h]
  \centering
  \includegraphics[width=16cm]{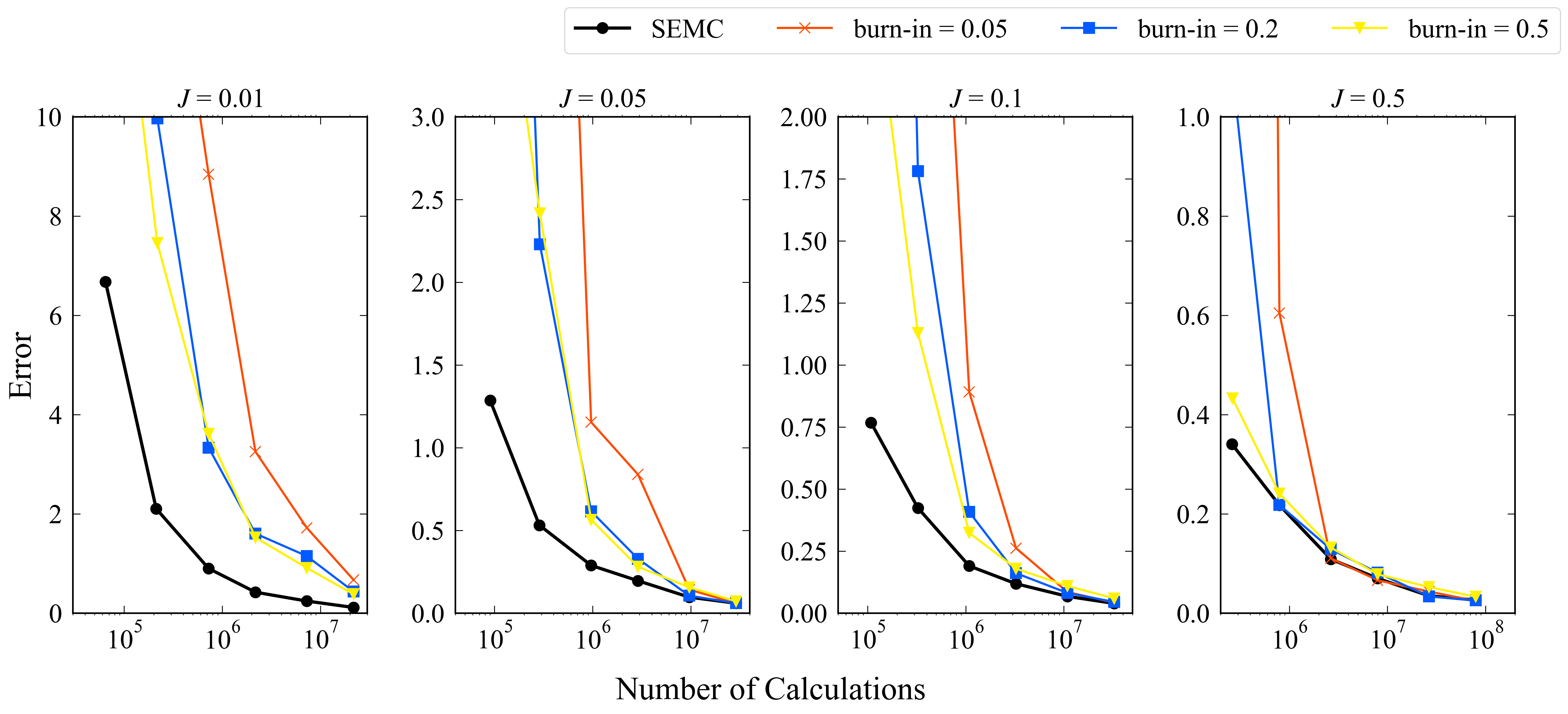}
  \caption{Comparison of SEMC and NRPT in estimating free energy across varying sample sizes and exchange rates.
  The mean absolute errors are shown for each method over 100 independent trials.
  From left to right, results correspond to exchange rates $J = 0.05, 0.1, 0.2, 0.5$.
  The black line represents SEMC, while the red, blue, and yellow lines represent NRPT with burn-in ratios of 5\%, 20\%, and 50\%, respectively.
  }
  \label{fig:multi_SEMC_REMC}
\end{figure}

This section compares the performance of SEMC and NRPT in estimating the free energy.
We fixed the correlation parameter at $r = 0$ and varied the number of samples across $600, 1800, 6000, 18000, 60000,$ and $180000$.
For each sample size, the target exchange rate $J$ was set to $0.05, 0.1, 0.2$, and $0.5$, and 100 independent runs of both SEMC and NRPT were conducted to compute the free energy.
Figure \ref{fig:multi_SEMC_REMC} shows the mean absolute error between the estimated and true free energies across these trials.
The results show that SEMC achieves higher accuracy than NRPT, particularly at lower exchange rates.

\begin{figure}[h]
  \centering
  \includegraphics[width=12cm]{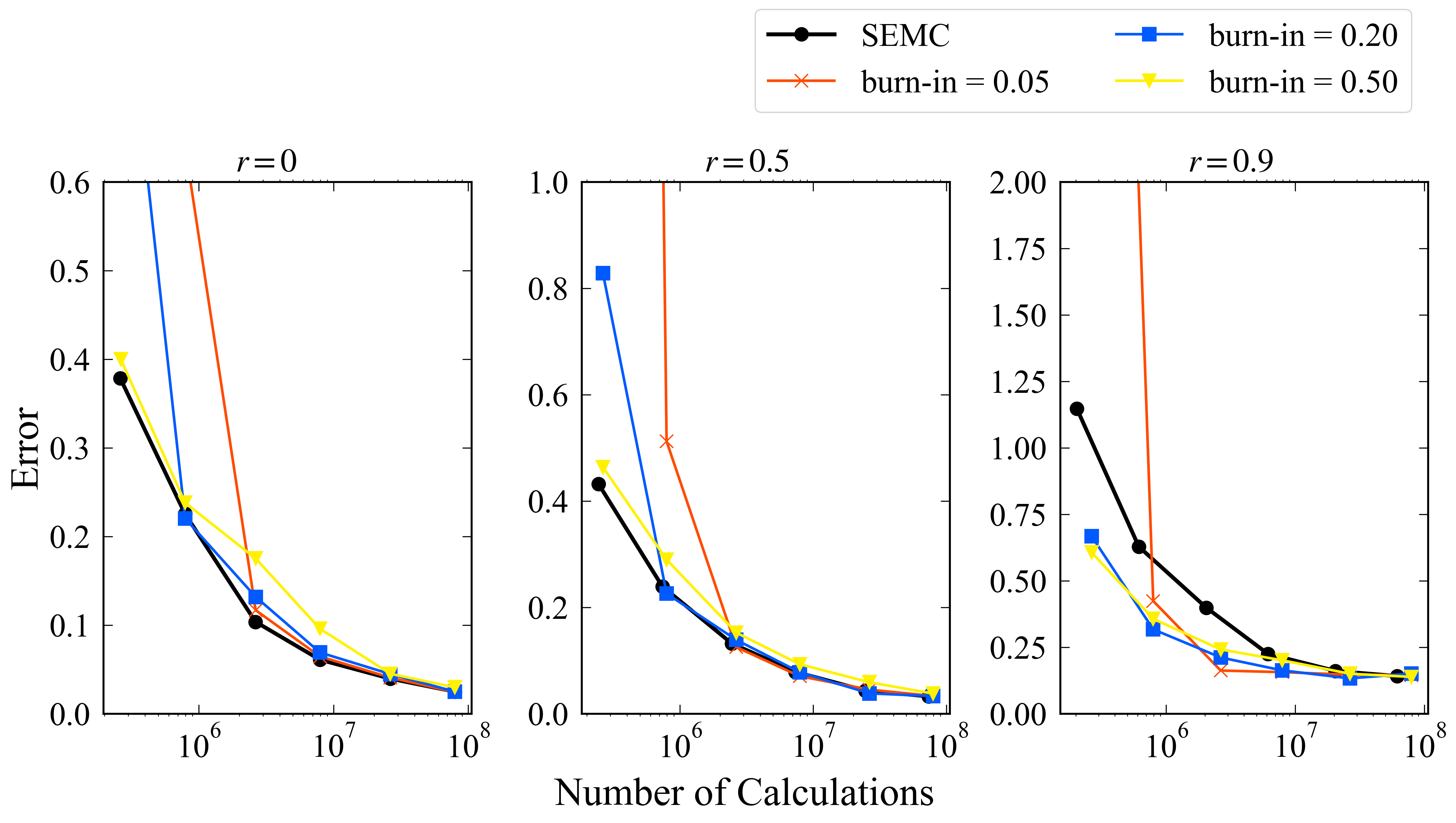}
  \caption{Comparison of SEMC and NRPT in estimating the free energy across varying correlation parameters $r$.
  The mean absolute errors are shown for each method over 100 independent trials.
  From left to right, the results correspond to $r = 0, 0.5, 0.9$.
  The black line represents SEMC, while the red, blue, and yellow lines represent NRPT with burn-in ratios of 5\%, 20\%, and 50\%, respectively.
  }
  \label{fig:multi_SEMC_REMC_r}
\end{figure}

We further evaluated performance at a fixed exchange rate, $J = 0.5$, by varying the correlation parameter $r = 0, 0.5, 0.9$.
Again, 100 runs of SEMC and NRPT were conducted to estimate the free energy, and the mean absolute errors are shown in Figure \ref{fig:multi_SEMC_REMC_r}.

When $r = 0$ or $r = 0.5$, SEMC consistently outperforms NRPT, with a burn-in ratio of 50\% in the large-sample regime.
In the small-sample regime, SEMC maintains higher accuracy than NRPT, even with shorter burn-in ratios of 5\% and 20\%.
However, for $r = 0.9$, NRPT yields more accurate estimates than SEMC.

\subsubsection{Comparison with SMCS}
\begin{figure}[h]
  \centering
  \includegraphics[width=16cm]{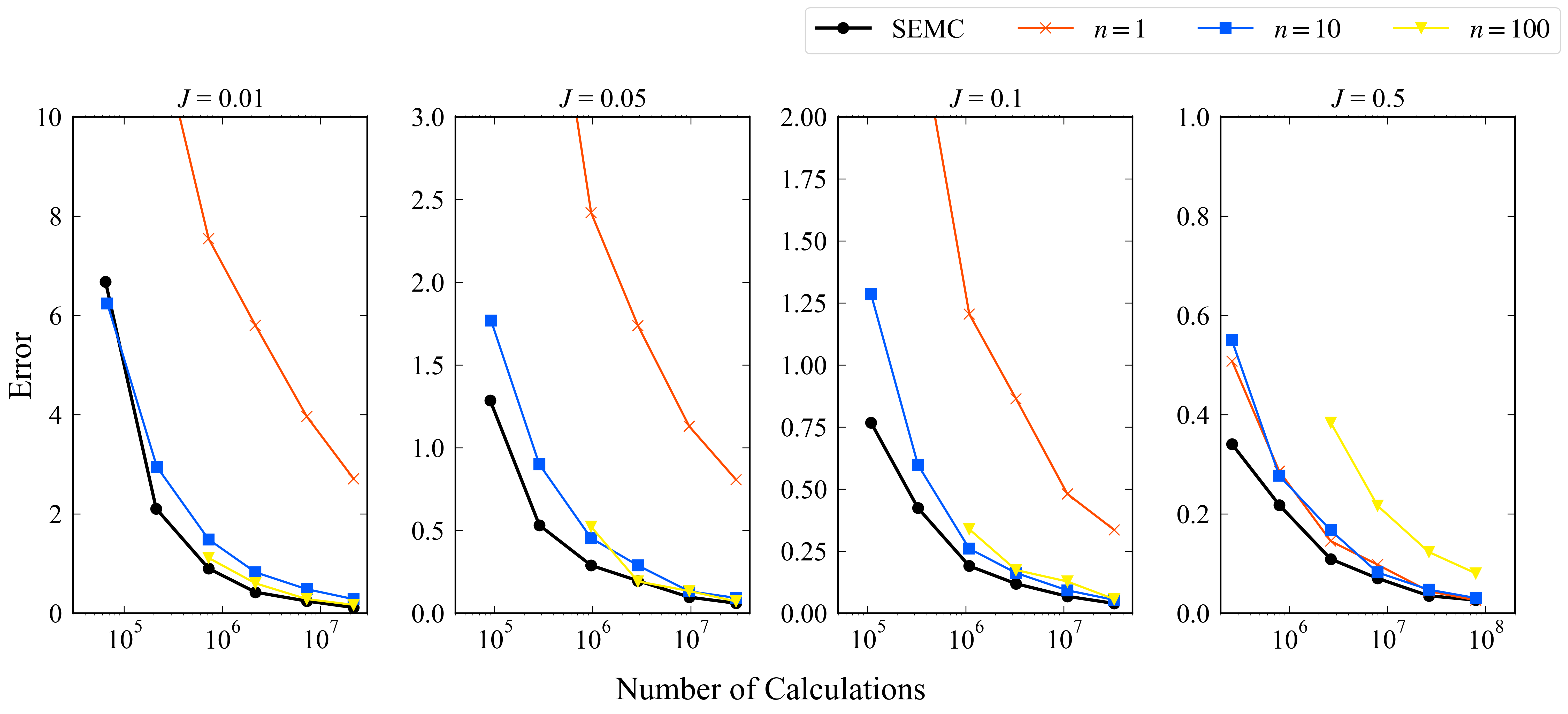}
  \caption{Comparison of SEMC and SMCS in estimating free energy across varying sample sizes and exchange rates.
  The mean absolute errors are shown for each method over 100 independent trials.
  From left to right, results correspond to exchange rates $J = 0.05, 0.1, 0.2, 0.5$.
  The black line represents SEMC, while the red, blue, and yellow lines represent SMCS with $n = 1, 10, 100$, respectively.
  }
  \label{fig:multi_SEMC_SMCS}
\end{figure}

We next compared SEMC with SMCS.
Using the same setting as in the NRPT comparison, we fixed $r = 0$ and varied the number of samples across $600$, $1800$, $6000$, $18000$, $60000$, and $180000$, with the target exchange rate $J = 0.05$, $0.1$, $0.2$, $0.5$.
For each configuration, 100 independent trials of SEMC and SMCS were conducted, and the mean absolute errors from the true free energy were computed, as shown in Figure \ref{fig:multi_SEMC_SMCS}.

The results indicate that, for SMCS, reducing the exchange rate $J$ necessitates increasing the number of MCMC steps $n$ per distribution to maintain accuracy.
In contrast, SEMC achieves higher accuracy without explicit tuning $n$, even at low exchange rates.

\begin{figure}[h]
  \centering
  \includegraphics[width=12cm]{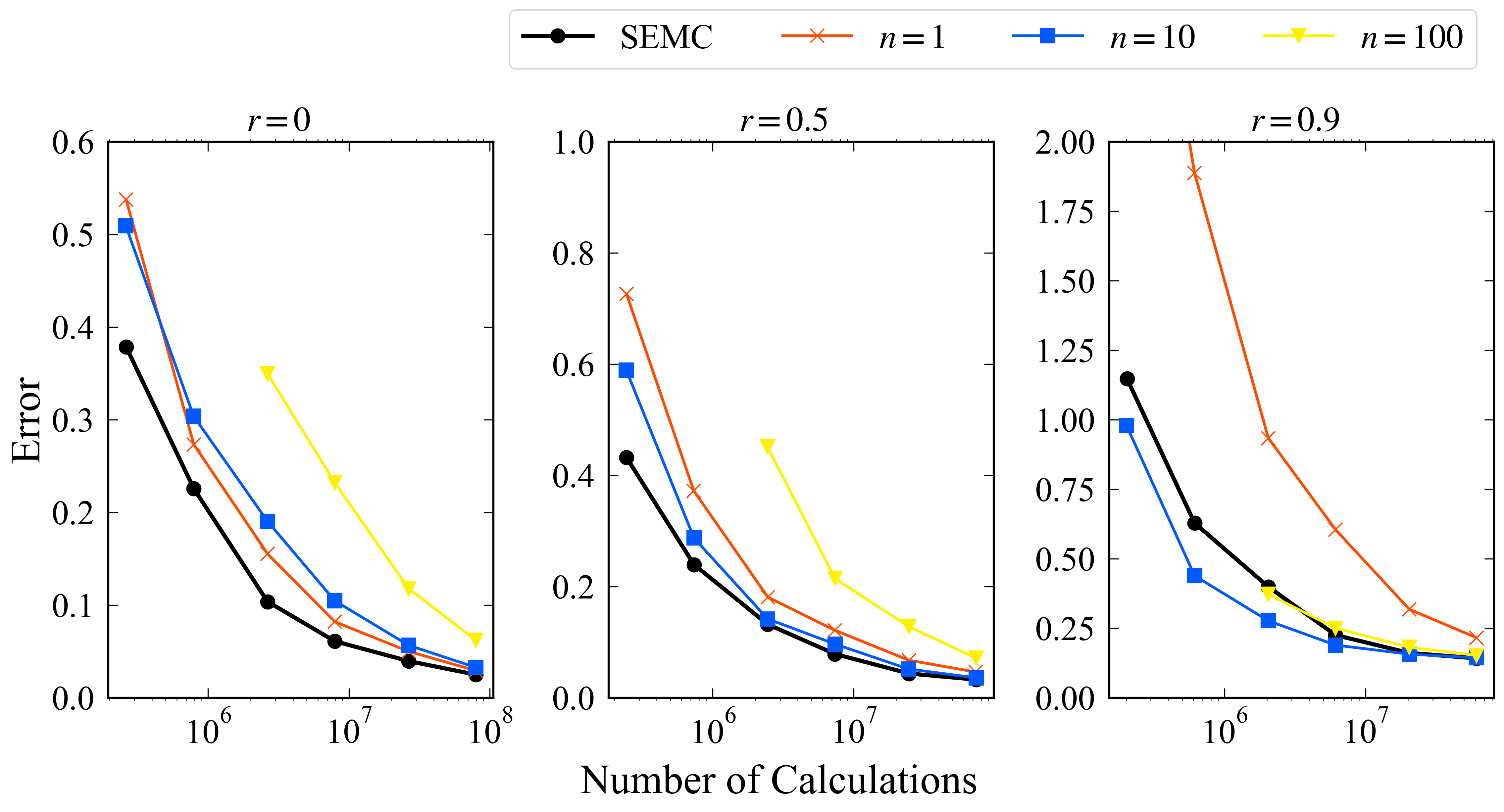}
  \caption{Comparison of SEMC and SMCS in estimating the free energy with varying correlation parameters $r$.
  The mean absolute errors are shown for each method across 100 independent trials.
  From left to right, results are shown for correlation parameters $r = 0, 0.5, 0.9$.
  The black line indicates estimated free energy using SEMC, and the red, blue, and yellow lines indicate estimated free energy using SMCS with $n = 1, 10, 100$, respectively.
  }
  \label{fig:multi_SEMC_SMCS_r}
\end{figure}

We also evaluated performance at a fixed $J = 0.5$ while varying $r = 0, 0.5, 0.9$, using the same experimental procedure as in Figure \ref{fig:multi_SEMC_SMCS_r}.
In SMCS, performance strongly depends on careful tuning of the number of MCMC steps $n$, particularly as the correlation increases.
In contrast, SEMC achieves higher accuracy than SMCS for $r = 0$ and $r = 0.5$, without requiring such parameter tuning.
However, similar to the NRPT comparison, SMCS with $n = 10$ outperforms SEMC when $r = 0.9$.

\subsection{Bayesian Inference in Materials Science}
This section applies SEMC to two Bayesian inference problems in materials science: spectral deconvolution and sparse linear regression.
\subsubsection{Problem Setting}
In spectral deconvolution, spectral data are modeled as the sum of multiple peak functions, and the number of peak functions and their parameters are estimated.
Let $\theta = \{a_k, \mu_k, b_k\}_{k=1}^K$ denote the parameters. The spectral function model $f_K(x;\theta)$ is defined as follows:
\begin{align}
  f_K(x;\theta) = \sum_{k=1}^K a_k\exp\left(-\frac{b_k}{2}(x - \mu_k)^2\right).
\end{align}
Assume that each observation $y_i$ follows a Gaussian distribution with mean $f_K(x_i;\theta)$ and variance $\sigma^2$.
The probability of obtaining the observed data, $D = \{x_i, y_i\}_{i=1}^N$, is as follows \cite{Nagata2012}:
\begin{align}
  p(D|\theta) = \prod_{i=1}^N \frac{1}{\sqrt{2\pi\sigma^2}}\exp\left(-\frac{(y_i - f_K(x_i;\theta))^2}{2\sigma^2}\right).
\end{align}
In this setting, we considered $K = 3$ and $K = 10$. 
The true parameter values and prior distributions for each $K$ are provided in Appendix\ref{sec:appendix_problem_setting}.


In sparse linear regression, the output is modeled as a linear combination of explanatory variables, with some coefficients constrained to zero.
Let $X \in \mathbb{R}^{N \times p}$ denote the input, $\boldsymbol{\beta} \in \mathbb{R}^{p}$ the coefficients, and $\boldsymbol{c} \in [0,1]^p$ an indicator vector determining whether each coefficient is included.
The output, $\boldsymbol{y} \in \mathbb{R}^{N}$, is then represented by the following linear regression model with noise $\boldsymbol{\epsilon} \in \mathbb{R}^{N}$:
\begin{align}
  \boldsymbol{y} = X(\boldsymbol{\beta} \odot \boldsymbol{c}) + \boldsymbol{\epsilon}.
\end{align}
Here, $\odot$ is the Hadamard product, namely $(\boldsymbol{\beta} \odot \boldsymbol{c})_i = \beta_i c_i$. \par
Assuming that each element of $\boldsymbol{\beta}$ follows an independent normal distribution with mean $0$ and variance $s$, and that $\boldsymbol{\epsilon}$ follows a normal distribution with mean $\boldsymbol{0}$ and covariance matrix $\Sigma$, the negative log-likelihood, $-\log(p(\boldsymbol{y}|\boldsymbol{c},X))$, is calculated as follows:
\begin{align}
  -\log(p(\boldsymbol{y}|\boldsymbol{c},X)) = K\log s + \frac{1}{2} \log\det(2\pi\Sigma) + \frac{1}{2}\log\det \Lambda - \frac{1}{2}\boldsymbol{\mu}^\top\Lambda^{-1}\boldsymbol{\mu} + \frac{1}{2}\boldsymbol{y}^\top\Sigma^{-1}\boldsymbol{y}.
  \label{eq:exhaustive}
\end{align}
Here, $K$ is the number of elements in $\boldsymbol{c}$ equal to $1$, $\Lambda = \left(X_I^\top \Sigma^{-1} X_I + \frac{1}{s^2}I\right)^{-1}$, $\boldsymbol{\mu} = \Lambda X_I^\top \Sigma^{-1} \boldsymbol{y}$, and $X_I$ is the matrix of explanatory variables corresponding to the nonzero elements of $\boldsymbol{c}$.
By minimizing the negative log-likelihood, the optimal subset of explanatory variables is selected.
When $K$ and $N$ are large, calculating Equation (\ref{eq:exhaustive}) for all combinations becomes infeasible.
Therefore, we define $E(\boldsymbol{c}) = -\log(p(\boldsymbol{y}|\boldsymbol{c},X))/N$ and employ the REMC, SMCS, and SEMC methods to find the optimal $\boldsymbol{c}$. 
In this setting, the Metropolis algorithm randomly selects an element of $\boldsymbol{c}$ and flips its value.
The true parameter values are provided in Appendix\ref{sec:appendix_problem_setting}.
Sparse linear regression is widely used for feature selection, including in materials science applications \cite{igarashi2018exhaustive}.
\par
\subsubsection{Tuning Parameters}
\begin{figure}[h]
  \centering
  \includegraphics[width = 16cm]{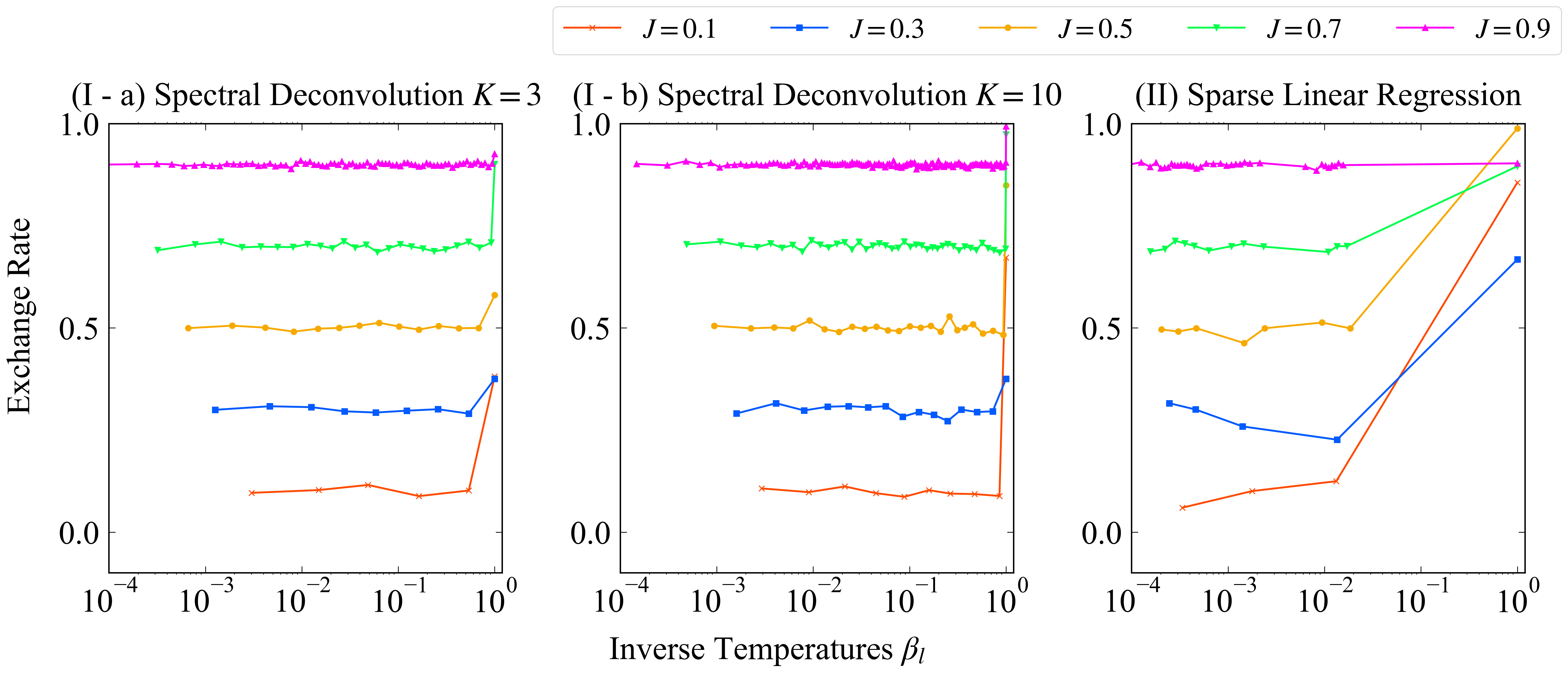}
  \caption{Inverse temperature, $(\beta_l)_{l=2}^L$, and exchange rate between the $p_{\beta_l}, p_{\beta_{l-1}}$ of the SEMC method with $J = 0.1, 0.3, 0.5, 0.7, 0.9$.
  From left to right, the results are shown for spectral deconvolution with $K = 3$ and $K = 10$, and sparse linear regression.
  }
  \label{fig:exchange_bayes_SEMC}
\end{figure}

This section presents the results of parameter tuning for SEMC.
We set the number of samples to 6000, and the target acceptance rate to $\alpha = 0.5$.
The target exchange rate $J$ was varied across $0.1$, $0.3$, $0.5$, $0.7$, and $0.9$, with the resulting actual exchange rates shown in Figure \ref{fig:exchange_bayes_SEMC}.
The results indicate that the exchange rates are accurately maintained across most temperatures, except between $\beta_L$ and $\beta_{L-1}$, as observed in Section 4.1.

\begin{figure}[h]
  \centering
  \includegraphics[width=12cm]{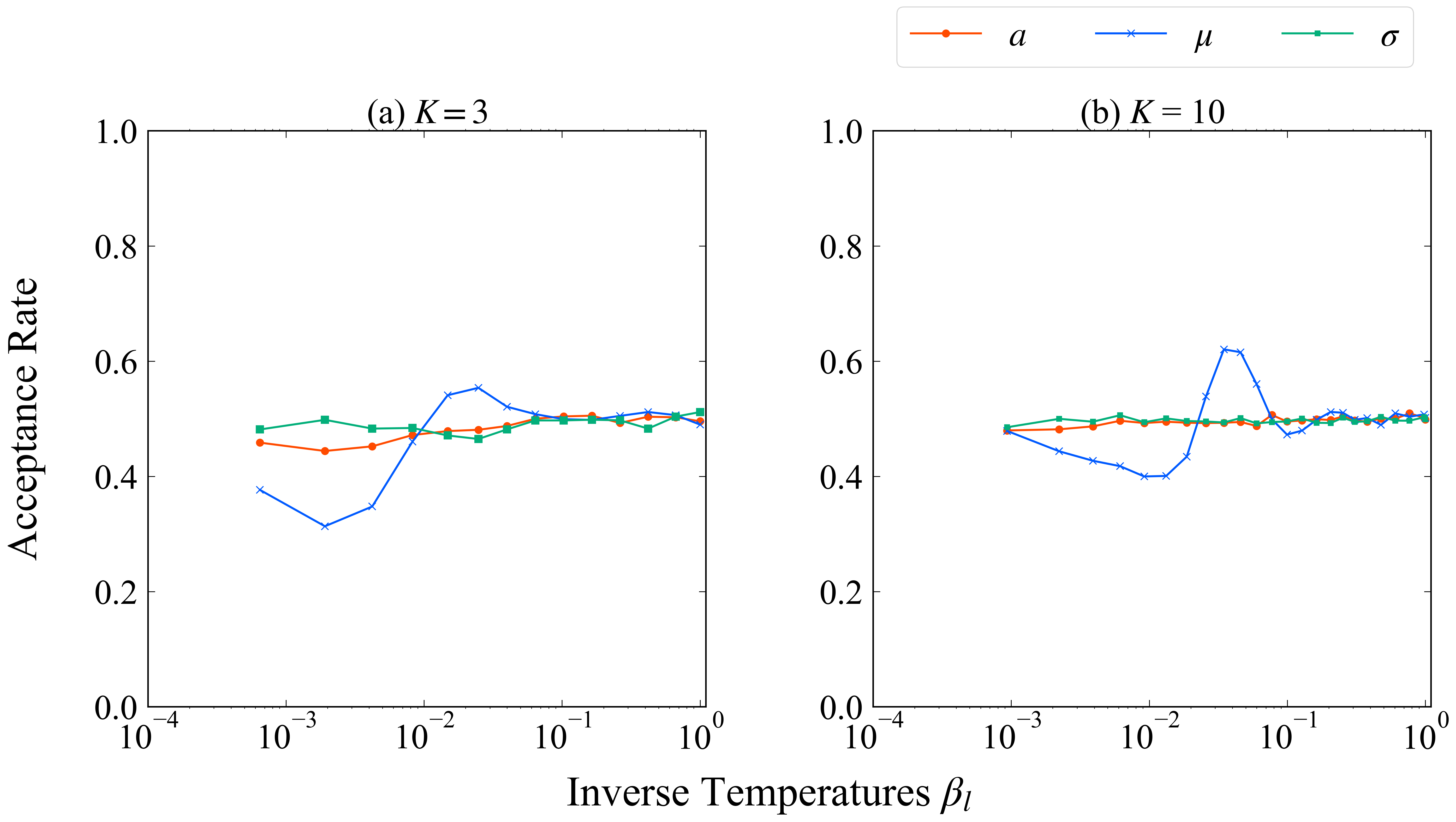}
  \caption{Acceptance rates of the Metropolis algorithm at each inverse temperature for SEMC with $J = 0.5$ and $r = 0$.
  The left and right panels show the results for spectral deconvolution with $K = 3$ and $K = 10$, respectively.
  The red, blue, and yellow lines indicate the acceptance rates for $\{a_k\}_{k=1}^K$, $\{\mu_k\}_{k=1}^K$, and $\{b_k\}_{k=1}^K$, respectively, in spectral deconvolution.
  }
  \label{fig:acceptance_bayes_SEMC}
\end{figure}

We also examined how well the Metropolis acceptance rates are controlled under the settings $\alpha = 0.5$ and $J = 0.5$, using 6000 samples and $r = 0$.
Figure \ref{fig:acceptance_bayes_SEMC} shows the average acceptance rates at each inverse temperature.
The results indicate that step-size tuning achieves the desired acceptance rates, particularly in regions with larger $\beta_l$, as observed in Section 4.1.

\subsubsection{Comparison with NRPT}
\begin{figure}[h]
  \centering
  \includegraphics[width=16cm]{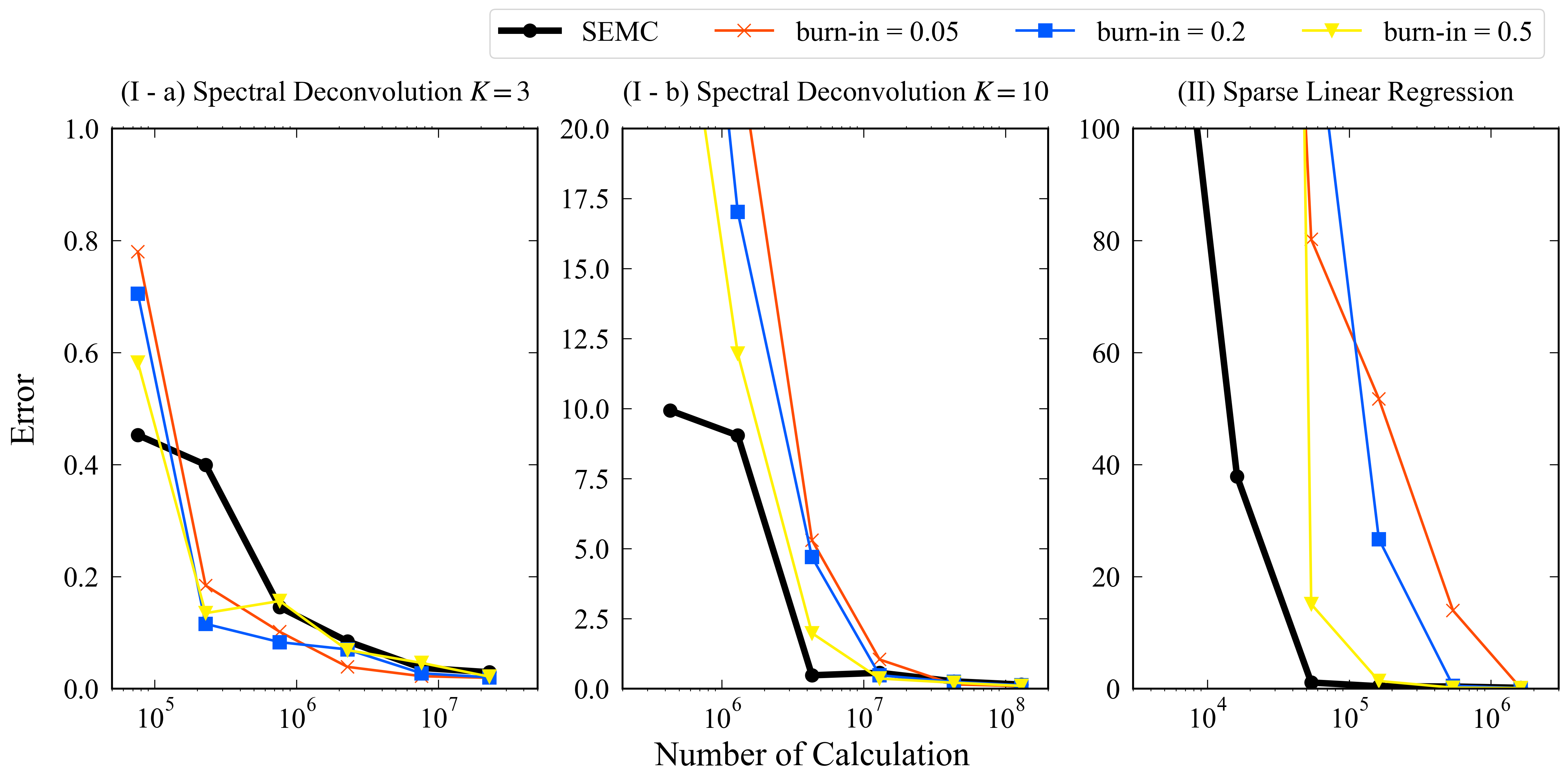}
  \caption{Comparison of SEMC and NRPT in estimating free energy across varying sample sizes.
  The mean absolute errors are shown for each method over 10 independent trials.
  From left to right, results correspond to spectral deconvolution with $K = 3$ and $K = 10$, and sparse linear regression.
  The black line represents SEMC, while the red, blue, and yellow lines represent NRPT with burn-in ratios of 5\%, 20\%, and 50\%, respectively.
  }
  \label{fig:bayes_SEMC_REMC}
\end{figure}

This section compares the performance of SEMC and NRPT in estimating the free energy.
The theoretical free energy was estimated by performing the NRPT method with a sufficient number of samples and intermediate distributions.
We set $J = 0.5$ and varied the number of samples across $600, 1800, 6000, 18000, 60000$, and $180000$.
For each sample size, 10 independent runs of both SEMC and NRPT were conducted to compute the free energy.
Figure \ref{fig:bayes_SEMC_REMC} shows the mean absolute error between the estimated free energy and the highly accurate reference value obtained from NRPT with a large number of samples.
The results indicate that SEMC achieves higher accuracy than NRPT in the small-sample regime. \par

\subsubsection{Comparison with SMCS}
\begin{figure}[h]
  \centering
  \includegraphics[width=16cm]{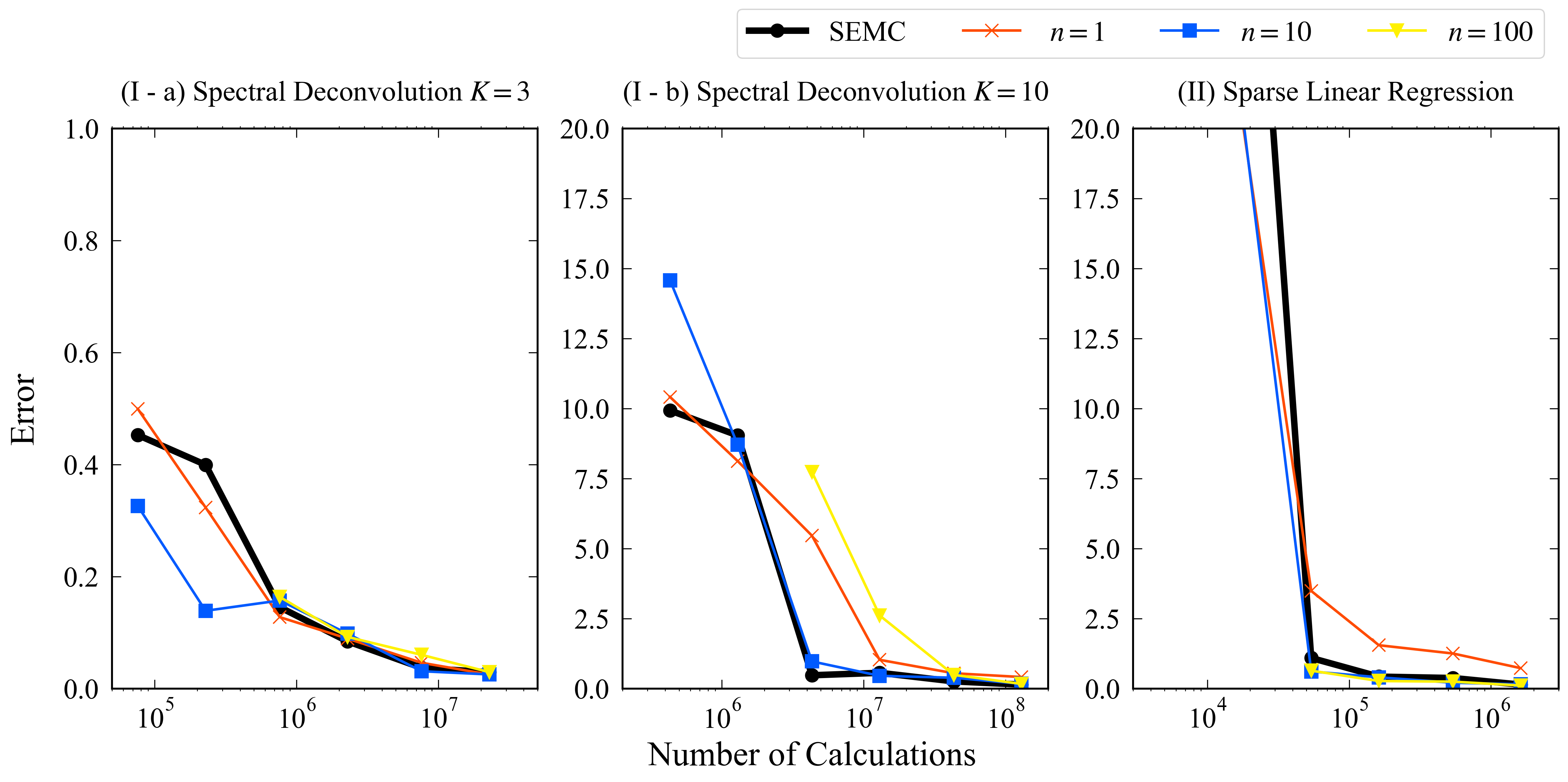}
  \caption{Comparison of SEMC and SMCS in estimating the free energy across varying sample sizes.
  The mean absolute errors are shown for each method over 10 independent trials.
  From left to right, results correspond to spectral deconvolution with $K = 3$ and $K = 10$, and sparse linear regression.
  The black line represents SEMC, while the red, blue, and yellow lines represent SMCS with $n = 1, 10, 100$, respectively.
  }
  \label{fig:bayes_SEMC_SMCS}
\end{figure}

We next compared SEMC with SMCS.
Using the same setting as in the NRPT comparison, we fixed $J = 0.5$ and varied the number of samples across $600$, $1800$, $6000$, $18000$, $60000$, and $180000$.
For each configuration, we conducted 10 independent trials of SEMC and SMCS, and the mean absolute errors from the true free energy were computed, as shown in Figure \ref{fig:bayes_SEMC_SMCS}.

In the spectral deconvolution problem, the results from SMCS and SEMC are similar.
On the other hand, in the sparse linear regression problem, SEMC outperforms SMCS with $n = 1$.

\section{Discussion}
This section discusses the properties of the SEMC method based on the results presented in Section 4. 
The comparison between SEMC and NRPT highlights the superior performance of SEMC in three main scenarios: (i) sampling from multimodal distributions with a small number of intermediate distributions, (ii) spectral deconvolution tasks with $K=10$, and (iii) sparse linear regression problems. 
These cases are characterized by large differences between intermediate distributions, which make convergence particularly challenging. 
Unlike NRPT, which initializes samples from the prior distribution, SEMC selects its initial samples from the previous distribution, accelerating convergence and improving performance. 
Moreover, whereas NRPT requires tuning the burn-in period to ensure convergence, SEMC eliminates the need for such tuning, providing a practical advantage.

Next, in the comparison between SEMC and SMCS, SEMC demonstrated comparable or superior performance across several key scenarios: (i) sampling from multimodal distributions with varying numbers of intermediate distributions, (ii) cases where the correlation parameter $r$ was 0 or 0.5 in the multimodal distributions, and (iii) sparse linear regression tasks, without requiring extensive parameter tuning. 
One reason for this performance difference is how the initial values of MCMC updates are handled during sampling. 
In SMCS, particularly under a small number of intermediate distributions, high-probability samples are often resampled repeatedly, and the Metropolis updates frequently start from identical points, resulting in redundant sampling paths and limited exploration. 
In contrast, SEMC rarely exchanges high-probability samples, and the Metropolis updates are performed sequentially from distinct samples, enabling more diverse sampling paths and better exploration of the sample space.
Although TMCMC, a variant of SMCS, updates resampled particles multiple times proportionally to the number of selections, such approaches have fallen out of use due to difficulties in accurately computing free energy \cite{wu2018bayesian}.
By incorporating “exchange” steps, SEMC preserves the theoretical validity of the sampling process while still allowing broader exploration around high-probability regions. 
Additionally, SEMC performs further sampling from the $(l-1)$-th distribution during the $l$-th distribution's sampling process via the exchange step, which likely contributes to its enhanced performance.

Nevertheless, when the correlation parameter is set to 
$r=0.9$, both NRPT and SMCS outperform SEMC. 
This decline in SEMC's performance is likely due to the strong parameter correlations, which reduce the effectiveness of the Metropolis updates. 
Consequently, the sampler fails to adequately explore the parameter space, resulting in a more biased empirical distribution.
To address this issue, enhancements to the MCMC kernel within SEMC, such as performing multiple Metropolis updates per sample, can improve sampling efficiency under strong correlations, as discussed in the Appendix \ref{sec:appendix_result_multiple_parameter_updates}.
This approach is conceptually analogous to increasing the number of MCMC steps in SMCS to maintain accuracy in complex settings.
In contrast, NRPT satisfies the detailed balance condition across the entire sampling process, making it less sensitive to the number of MCMC steps. 
This property enables NRPT to maintain reliable performance without additional tuning, representing a practical advantage over SEMC and SMCS in highly correlated scenarios.

\section{Conclusion and Future Work}

In this study, we comparatively investigated sampling methods that enable Bayesian data analysis for complex models with little manual tuning, including NRPT, SMCS, and our newly proposed SEMC, which integrates exchange steps with SMCS to sample from multimodal distributions.
Numerical experiments demonstrated that SEMC outperforms NRPT, achieving higher accuracy with fewer samples. 
Moreover, unlike SMCS, SEMC maintains its performance without requiring tuning of the MCMC steps, particularly in multimodal distributions with a small number of intermediate distributions and in sparse linear regression problems.

In future work, we plan to explore alternative MCMC kernels to replace the Metropolis algorithm.
Our SEMC experiments suggested that, when parameter correlations are high and the MCMC kernel performs poorly, the accuracy of free energy estimation declines. 
To address this, the parameter-update stage could adopt techniques such as Langevin dynamics, Hamiltonian Monte Carlo, as used in SMCS, or more recent approaches, including multiple-try Metropolis \cite{liu2000multiple} and piecewise deterministic Markov processes \cite{bierkens2019zig}. 
Additionally, SEMC allows flexible spacing of intermediate distributions, which could be exploited to design distributions that are optimal for computing Bayesian free energy and related quantities.

\section*{Acknowledgment}
This work was supported by JSPS KAKENHI Grant Number JP25KJ0959.

\appendix
\section{Detailed Algorithms}
\label{sec:appendix_algorithm}
This section describes the detailed algorithms for the three methods used in this study.

  \begin{algorithm}[h]
    \caption{Replica Exchange Monte Carlo}
    \label{EMC}
    \begin{algorithmic}[1]
    \REQUIRE Data $D$, model $M$, prior distribution $p(\theta|M)$, set of inverse temperatures $(\beta_l)_{l = 1}^L$, burn-in ratio $\rho$, sample size $T$, step size $(\epsilon_l)_{l = 2}^L$
    \ENSURE Parameter set $\Theta^{T}$ = $\left( \{\theta_l^t\}_{t = 1}^{T} \mid \theta_l^t \textup{ follows } p_{\beta_l}(\theta) \right)_{l = 1}^L$
    \STATE Parameter set $\Theta^{T}$ = \{\}
    \FOR{$l \in \{1,...,L\}$}
        \STATE Generate $\theta_{l}$ according to the prior distribution $p(\theta|M)$.
    \ENDFOR
    \FOR{$t \in (1,...,T)$}
      \STATE Generate $\theta_{1}$ according to the prior distribution $p(\theta|M)$.
      \FOR{$l \in \{2,...,L\}$}
        \STATE Update $\theta_{l}$ according to $p_{\beta_l}(\theta)$ using the Metropolis algorithm with step size $\epsilon_l$.
      \ENDFOR
      \STATE Choose $l$ and $l+1$ according to the exchange scheme.
      \STATE Calculate $v = \frac{p_{\beta_{l}}(\theta_{l+1})p_{\beta_{l+1}}(\theta_{l})}{p_{\beta_{l}}(\theta_{l})p_{\beta_{l+1}}(\theta_{l+1})}$.
      \STATE Exchange $\theta_l$ and $\theta_{l+1}$ with probability $\min(1,v)$.
      \IF{$t > T \times \rho$}
        \STATE Add $(\theta_l)_{l=1}^L$ to $\Theta$.
      \ENDIF
    \ENDFOR
    \end{algorithmic}
  \end{algorithm}


\begin{algorithm}[h]
  \caption{Waste-Free Sequential Monte Carlo Samplers}
  \label{Waste-free SMC}
  \begin{algorithmic}[1]
    \REQUIRE Data $D$, prior distribution $p(\theta|M)$, sample size $T$, number of steps $n$, number of samples $S = T/n$, step size matrix $(\epsilon_l)_{l = 2}^L$
    \ENSURE Parameter set $\Theta$ = $\left((\theta_l^t)_{l = 1}^L \mid \theta_l^t \textup{ follows } p_{\beta_l}(\theta)\right)_{t = 1}^{T}$
    \STATE Parameter set $\Theta$ = ()
    \FOR{$i \in \{1,...,T\}$}
        \STATE Generate $\theta_1^i$ according to the prior distribution $p(\theta|M)$.
    \ENDFOR
    \STATE Add $(\theta_1^i)_{i=1}^T$ to $\Theta$.
    \STATE Set $l = 1, \beta_1 = 0$.
    \WHILE{$\beta_l < 1$}
        \STATE Set the next temperature $\beta_{l+1}$ and $l = l + 1$.
        \STATE Calculate $W_l = \{ \exp(-(\beta_{l} - \beta_{l-1})NE(\theta_{l-1}^i))\}_{i=1}^T$.
        \FOR{$i \in \{1,...,S\}$}
            \STATE Select $j$ randomly such that $\theta_l^i = \theta_{l-1}^j$ with probability $\frac{W_l^j}{\sum_j W_l^j}$.
        \ENDFOR
        \FOR{$j \in (1,...,n-1)$}
            \FOR{$i \in \{1,...,S\}$}
                \STATE Set $\theta_{l}^{S\times j+i} = \theta_{l}^{1,S\times (j-1)+i}$. 
                \STATE Update $\theta_{l}^{S\times j+i}$ using the MCMC algorithm with step size $\epsilon_l$.
            \ENDFOR
        \ENDFOR
        \STATE Add $(\theta_l^i)_{i=1}^{T}$ to $\Theta$.
    \ENDWHILE
  \end{algorithmic}
\end{algorithm}
\newpage


\begin{algorithm}[h]
  \caption{Sequential Exchange Monte Carlo}
  \label{SEMC}
  \begin{algorithmic}[1]
    \REQUIRE Data $D$, prior distribution $p(\theta|M)$, set of inverse temperatures $(\beta_l)_{l = 1}^L$, sample size $T$, step size $(\epsilon_l)_{l = 2}^L$
    \ENSURE Parameter set $\Theta$ = $\left((\theta_l^t)_{l = 1}^L \mid \theta_l^t \sim p_{\beta_l}(\theta)\right)_{t = 1}^{T}$
    \STATE Parameter set $\Theta$ = ()
    \FOR{$i \in \{1,...,T\}$}
        \STATE Generate $\theta_1^i$ from the prior distribution $p(\theta|M)$.
    \ENDFOR
    \STATE Add $(\theta_1^i)_{i=1}^T$ to $\Theta$.
    \FOR{$l \in (2,...,L)$}
        \STATE Calculate $W_l = \{ \exp(-(\beta_{l} - \beta_{l-1})NE(\theta_{l-1}^i))\}_{i=1}^{T}$.
        \STATE Select $j$ such that $\theta_{l}^{1} = \theta_{l-1}^j$ with probability $\frac{W_l^j}{\sum_j W_l^j}$.
        \FOR{$i \in (2,...,T)$}
            \STATE Set $\theta_l^i = \theta_{l}^{i-1}$.
            \STATE Update $\theta_l^i$ using the Metropolis algorithm with step size $\epsilon_l$.
            \STATE Select $j$ from $\{1,...,T\}$ uniformly at random.
            \STATE Calculate $v = \dfrac{p_{\beta_{l}}(\theta_{l-1}^j)p_{\beta_{l-1}}(\theta_l^i)}{p_{\beta_{l}}(\theta_l^i)p_{\beta_{l-1}}(\theta_{l-1}^j)}$.
            \STATE Exchange $\theta_{l-1}^j$ and $\theta_l^i$ with probability $\min(1,v)$.
        \ENDFOR
        \STATE Add $(\theta_l^i)_{i=1}^{T}$ to $\Theta$.
    \ENDFOR
  \end{algorithmic}
\end{algorithm}
\newpage
\newpage
\section{Parameter Tuning for REMC}
\label{sec:appendix_tuning}

\subsection{Tuning of Step Size}
As described in the main text, for the REMC method, we tuned the step size of the Metropolis algorithm to achieve an acceptance rate of approximately 0.5 at each temperature.
Step-size tuning was performed using the Robbins-Monro method  \cite{robbins1951stochastic, garthwaite2016adaptive} and the dual averaging method \cite{nesterov2009primal, hoffman2014no}.
 In the Robbins-Monro method, the step size, $\epsilon$, is updated according to the following equation to ensure that the current acceptance rate, $p_{\textup{accept}}$, converges to the target $p^*$ during the burn-in period:
\begin{align}
  \epsilon^{\textup{new}} = \epsilon^{\textup{old}} + \epsilon^{\textup{old}} \times c \times \frac{p_{\textup{accept}} - p^*}{N_0 + N}.
\end{align}
Here, $c, N_0$ were set to appropriate values. This study updated the step size every $M = 20$ samples and set $c = 4, N_0 = 15$, and $p^* = 0.5$ for all distributions and parameters. \par

The dual averaging method updates the step size, $\epsilon$, using the following update equation:
\begin{align}
  \log \epsilon^{\textup{new}} &= (1-\eta) \log \epsilon^{\textup{old}} + \eta \log \widetilde{\epsilon}, \\
  H_t &= \left(1 - \frac{1}{t + t_0}\right) H_{t-1} + \frac{1}{t + t_0} (p_{\textup{accept}} - p^*), \\
  \log \widetilde{\epsilon} &= \mu - \frac{\sqrt{t}}{\gamma} \sum_{i=1}^{t} H_t. \\
\end{align}
Here, we set $\eta = t^{-\kappa}, \kappa = 0.75, t_0 = 10, \gamma = 0.05, H_0 = 0,$ and $\mu = \log(10\epsilon^{initial}),$ as in \cite{hoffman2014no}. \par

For both methods, the initial step size should be set to a value that achieves an acceptance rate of approximately $p^*$.
In this study, we set the initial step size from the following equation proposed by \cite{nagata2024algebraic}:

\begin{align}
U_{l,i} \sim \frac{(\log N\beta_l)^{m_i - m}}{(N\beta_l)^{\lambda_i - \lambda}} \cdot \frac{c}{\sigma_{l,i}},
\label{eq:acceptance_rate_appendix}
\end{align}

where
\begin{itemize}
  \item $\lambda$ and $m$ are the pole and its order of the zeta function $\zeta(z) = \int E(\theta)^z p(\theta) \, d\theta$,
  \item $\lambda_i$ and $m_i$ are those of the zeta function $\zeta_i(z) = \int |\theta_i| E(\theta)^z p(\theta) \, d\theta$,
  \item $\sigma_{l,i}$ is the step size of the proposal distribution for $\theta_{l,i}$,
  \item $c$ is a constant depending on $E(\theta)$ and $p(\theta)$.
\end{itemize}

For the regular model, $m = m_i = 1$ and $\lambda = \frac{d}{2}, \lambda_i = \frac{d+1}{2}$, where $d$ is the dimension of the parameter space.
Therefore, we set the initial step size as follows:
\begin{align}
  \epsilon_{l}^{initial} = \left( \frac{\beta_l}{\beta_1}\right)^{-\frac{1}{2}} \epsilon_1^{initial}.
\end{align}

We set $\epsilon_1^{initial}$ so that the acceptance rate of the Metropolis algorithm is approximately 0.5 at $\beta_1 = 0$.

Here, we describe the step size $\epsilon_1^{initial}$ that yields an approximate acceptance rate of 0.5 for the Metropolis algorithm when the target distribution is uniform.
First, we note that the theoretical acceptance rate for a uniform distribution is 0.5 when the step size equals the width of the distribution.
Let $p(\theta)$ be the uniform distribution on $[a,b]$,
and $\epsilon$ be the step size.
The acceptance rate of the Metropolis algorithm is then given by
\begin{align}
    R &= \int \int_{-\epsilon}^{\epsilon} \min\left(1,\frac{p(\theta + \delta)}{p(\theta)}\right)p(\delta)p(\theta)d\delta d\theta \\
    &= \frac{1}{b-a}\frac{1}{2\epsilon} \int_a^b \int_{-\epsilon}^{\epsilon} 1_{[a,b]}(\theta + \delta)d\delta d\theta \\
    &= \frac{1}{b-a}\frac{1}{2\epsilon} \int_a^b \int_{-\epsilon}^{\epsilon} 1_{[a-\theta,b-\theta]}(\delta)d\delta d\theta,
\end{align}
where $1_{[a,b]}$ is the indicator function of $[a,b]$ when $\epsilon = b-a$, $ [a-\theta,b-\theta] \subset [-\epsilon,\epsilon]$ for all $\theta \in [a,b]$.
Thus, $\int_{-\epsilon}^{\epsilon} 1_{[a-\theta,b-\theta]}(\delta)d\delta = b-a = \epsilon$, and $R = \frac{1}{b-a}\frac{1}{2\epsilon} \int_a^b \int_{-\epsilon}^{\epsilon} 1_{[a-\theta,b-\theta]}(\delta)d\delta d\theta = 0.5$.

\begin{figure}[h]
    \centering
    \includegraphics[width=8cm]{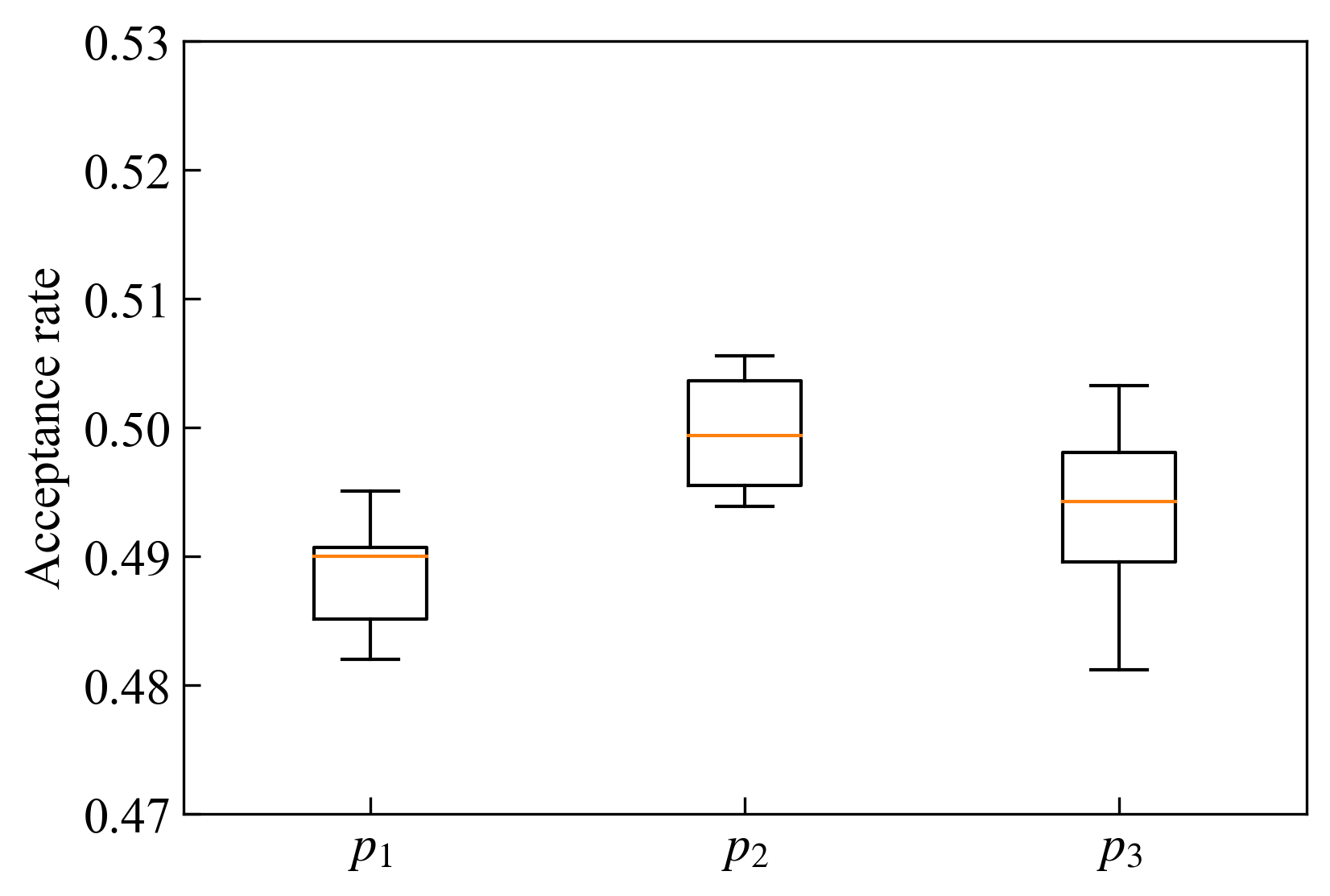}
    \caption{Acceptance rate of Metropolis-Hastings algorithm for various distributions. The vertical and horizontal axes show the acceptance rate and distribution, respectively.}
    \label{fig:acceptance_rate}
\end{figure}

Next, we show that the acceptance rate for the other distributions used in this study was approximately 0.5 when the step size was 2.94 times the variance of the distribution by simulations.
Let $\eta_a = 5.0, \lambda_a = 5.0, \nu_0 = 1.5, \xi_0 = 5.0, \eta_\sigma = 5.0$, and $\lambda_\sigma = 0.04$. We defined the distribution, $p_1(\theta),p_2(\theta),p_3(\theta)$ as
\begin{align}
    p_1(\theta) &\propto \textup{Gamma} \left(\exp(\theta);\eta_a,\lambda_a\right) \\
    p_2(\theta) &= N(\theta;\nu_0, (\xi_0)^{-1}) \\
    p_3(\theta) &\propto \textup{Gamma} \left(\exp(\theta);\eta_{\sigma},\lambda_{\sigma}\right)
\end{align}
Let the standard deviations of the distributions be $\sigma_1,\sigma_2,\sigma_3$. We set the step sizes, $\epsilon_1, \epsilon_2, \epsilon_3$ to $2.94\sigma_1,2.94\sigma_2,2.94\sigma_3$, respectively.
The acceptance rates of the Metropolis algorithm were estimated via simulation.
For each distribution, 10{,}000 samples were generated using the Metropolis algorithm with the corresponding step size, and the simulation was repeated 10 times.
Figure \ref{fig:acceptance_rate} shows the resulting acceptance rates, which were approximately 0.5 for all distributions.

\subsection{Results of Tuning the Step Size}
\begin{figure}[h]
  \centering
  \includegraphics[width=12cm]{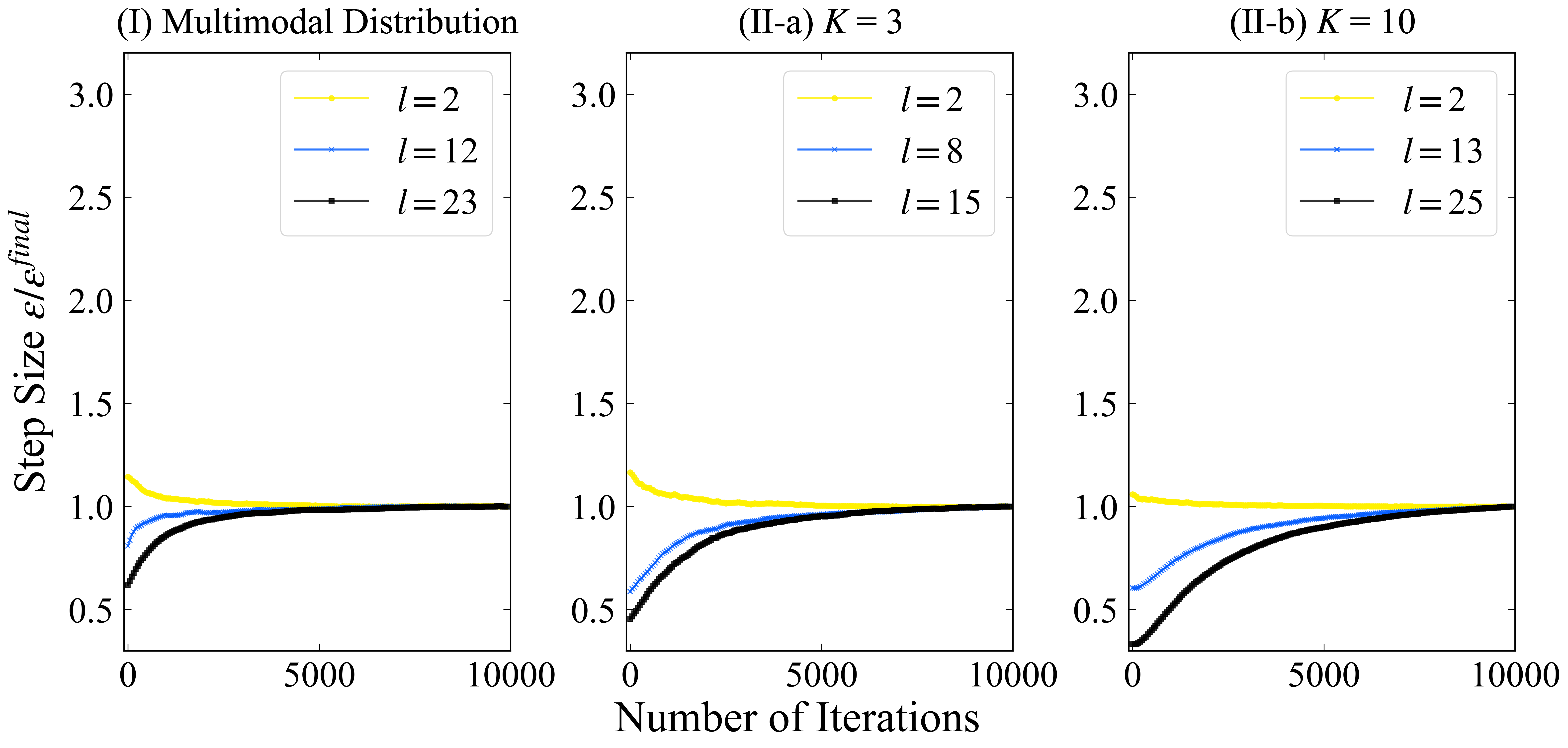}
  \caption{Transition of step size during the burn-in period using the Robbins-Monro method. (I) Sampling from a multimodal distribution with $r = 0$; (II-(a)) spectral deconvolution ($K=3$); (II-(b)) spectral deconvolution ($K=10$). 
  The high, intermediate, and low temperatures are indicated by round, cross, and square markers, respectively. 
  The vertical and horizontal axes represent the ratio of the step size to the final step size, $\epsilon/\epsilon^{final}$, and the number of iterations, respectively.}
  \label{fig:step_size_robbin}
\end{figure}

\begin{figure}[h]
  \centering
  \includegraphics[width=12cm]{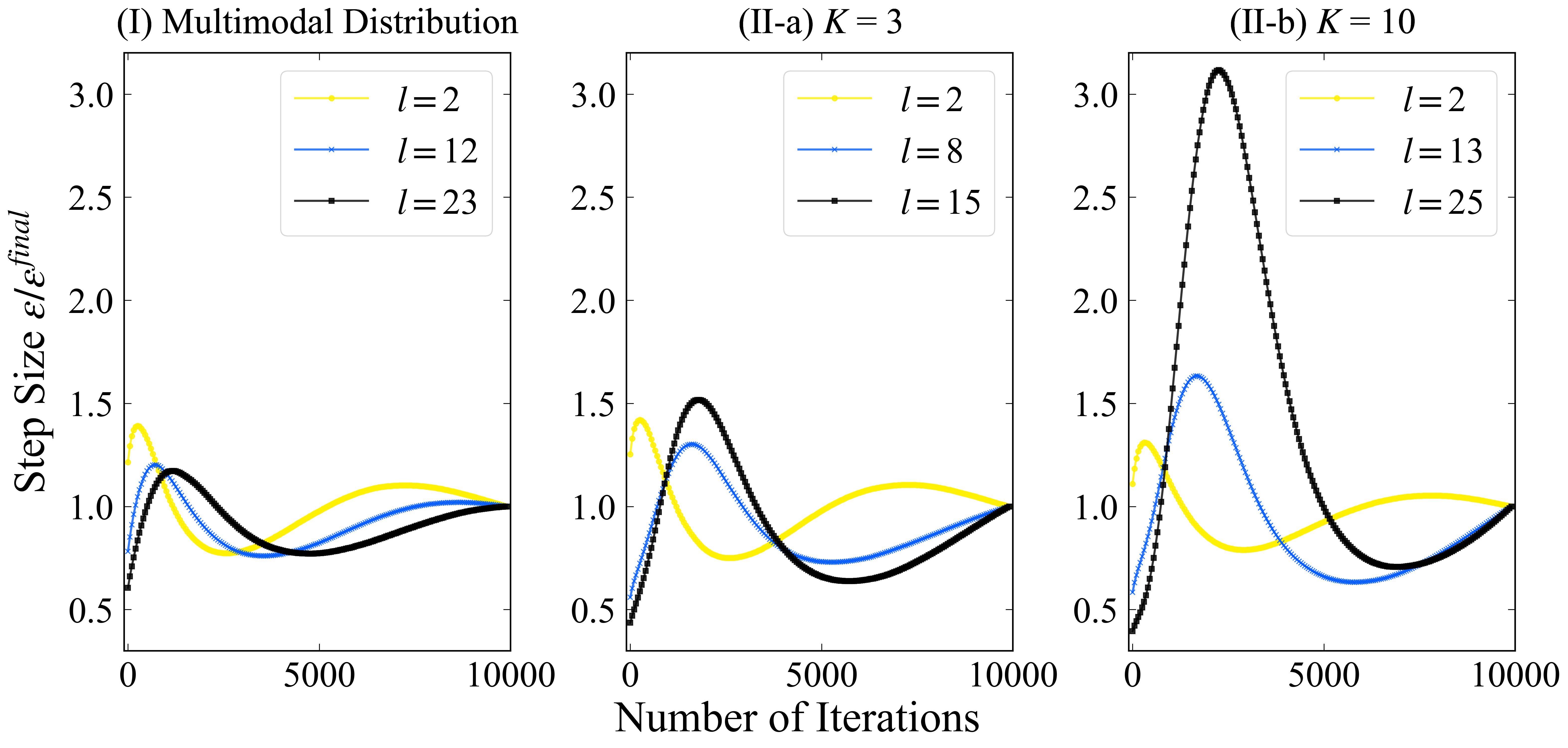}
  \caption{Transition of step size during the burn-in period using the dual averaging method. (I) Sampling from a multimodal distribution with $r = 0$; (II-(a)) spectral deconvolution ($K=3$); (II-(b)) spectral deconvolution ($K=10$). 
  High, intermediate, and low temperatures are indicated by round, cross, and square markers, respectively. 
  The vertical axis shows the ratio of the step size to the final step size, $\epsilon/\epsilon^{final}$, and the horizontal axis shows the number of iterations.}
  \label{fig:step_size_dual}
\end{figure}

We compared the performance of the Robbins-Monro method and the dual averaging method.
Let $\epsilon^{final}$ denote the final step size.
We examined the evolution of $\epsilon/\epsilon^{final}$ during the burn-in period for the SEMC method.
Figure \ref{fig:step_size_robbin} and \ref{fig:step_size_dual} show this evolution of $\epsilon/\epsilon^{final}$ at a high temperature $(l = 2)$, an intermediate temperature $(l = \lfloor (L+2)/2 \rfloor)$, and a low temperature $(l = L)$ in the continuous-variable problem setting.
The plotted values represent averages over all parameters. The Robbins-Monro method converged to the final step size more steadily than the dual averaging method.
\newpage
\newpage

\begin{figure}[h] 
  \centering
  \includegraphics[width=16cm]{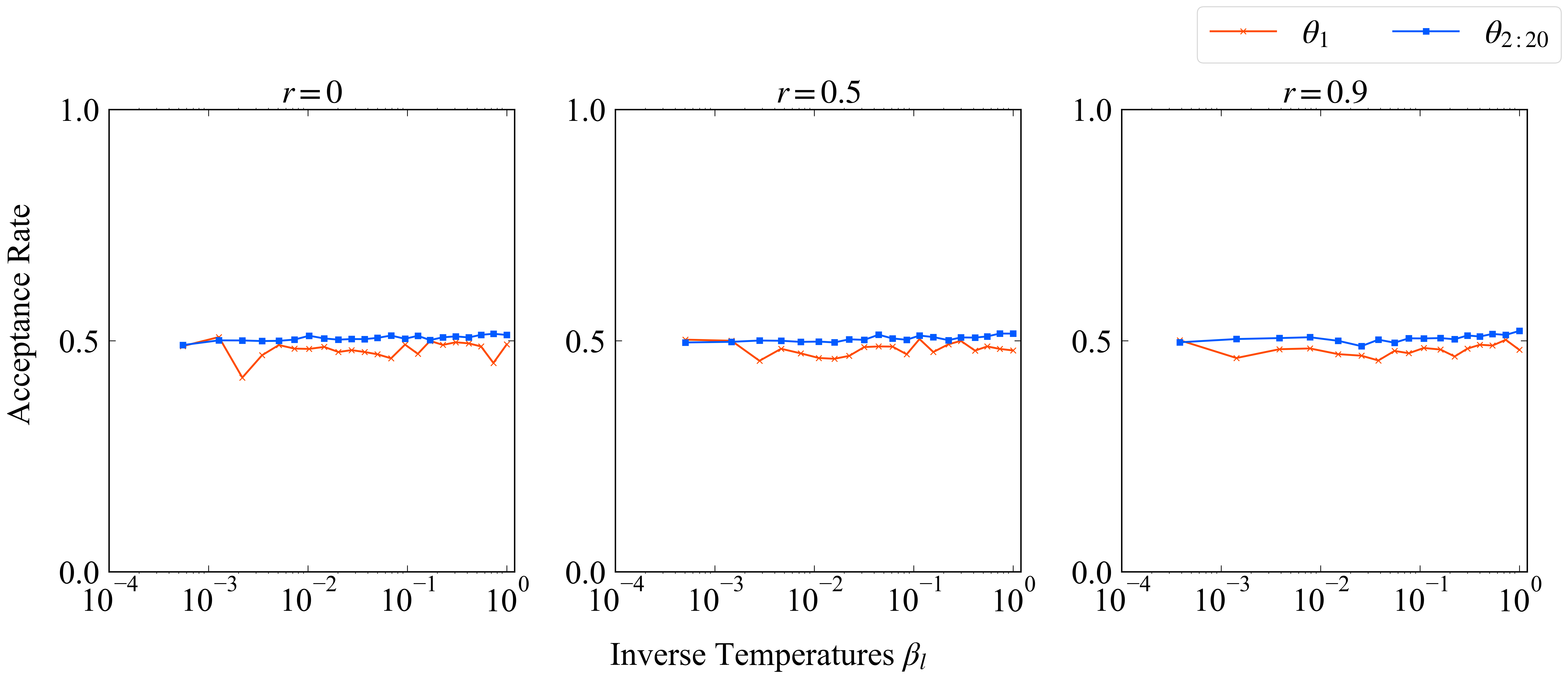}
  \caption{Acceptance rates of the Metropolis algorithm at each inverse temperature level for the multimodal distribution sampling with $J = 0.5$.
  From left to right, results correspond to correlation parameters $r = 0, 0.5, 0.9$.
  The red line represents the acceptance rate for $\theta_1$, and the blue line represents the average acceptance rate for $\theta_2, \ldots, \theta_d$.}
  \label{fig:acceptance_multi_REMC}
\end{figure}

\begin{figure}[h]
  \centering
  \includegraphics[width=12cm]{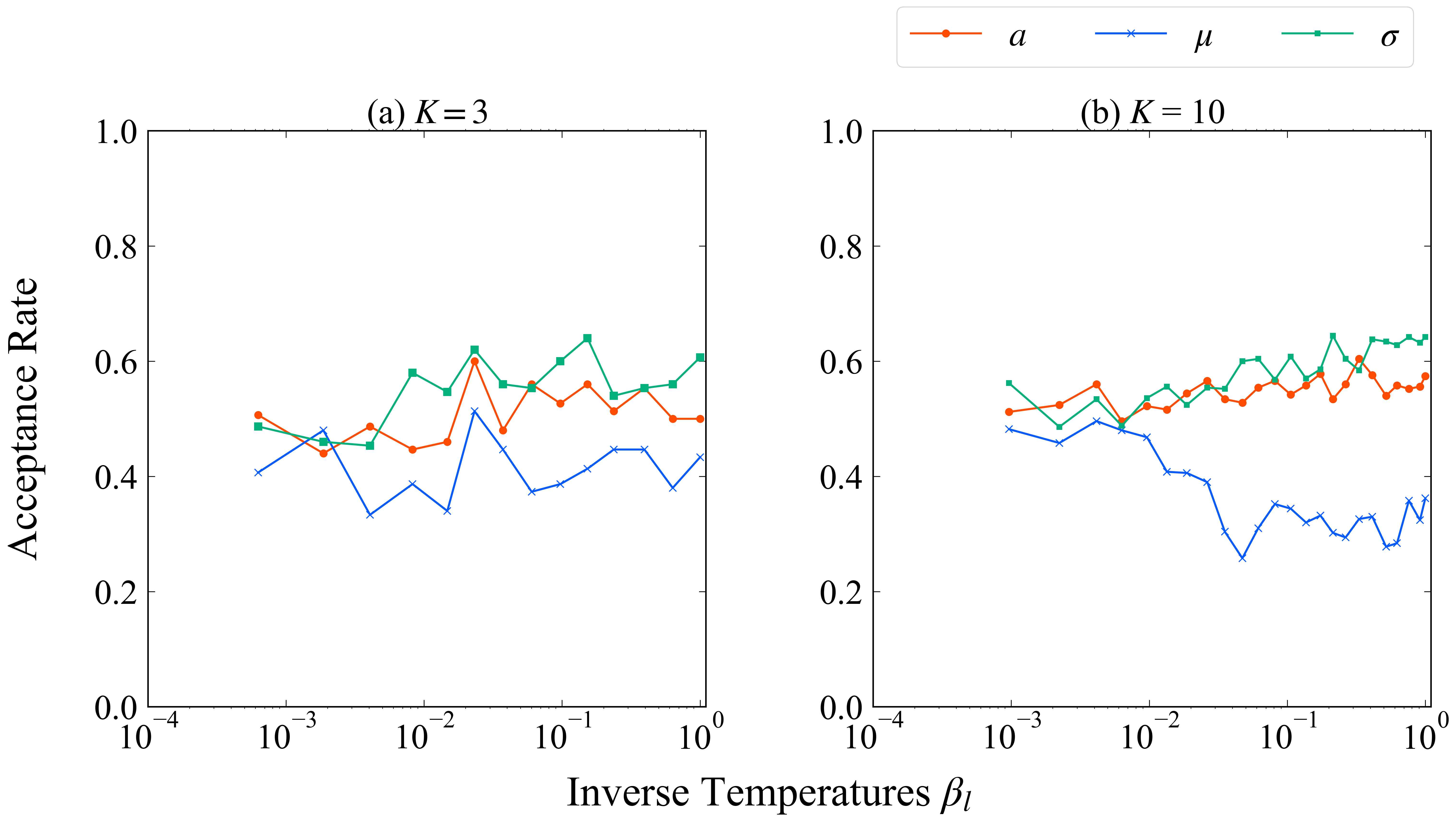}
  \caption{Acceptance rates of the Metropolis algorithm at each inverse temperature for Bayesian spectral deconvolution with $J = 0.5$.
  The left and right panels correspond to spectral deconvolution with $K = 3$ and $K = 10$, respectively.
  The red, blue, and yellow lines indicate the acceptance rates for $\{a_k\}_{k=1}^K$, $\{\mu_k\}_{k=1}^K$, and $\{b_k\}_{k=1}^K$, respectively.}
  \label{fig:acceptance_bayes_REMC}
\end{figure}

Next, we compared the acceptance rates of the Metropolis algorithm across temperatures.
Figure \ref{fig:acceptance_multi_REMC} and \ref{fig:acceptance_bayes_REMC} show the average acceptance rates at each temperature with the Robbins-Monro method.
The sample size was set to 6000, with a burn-in period of 50\% of the samples.
The results indicate that the acceptance rate remained around 0.5 across all temperatures.

\subsection{Result of Tuning the Temperatures}

\begin{figure}[h]
  \centering
  \includegraphics[width=16cm]{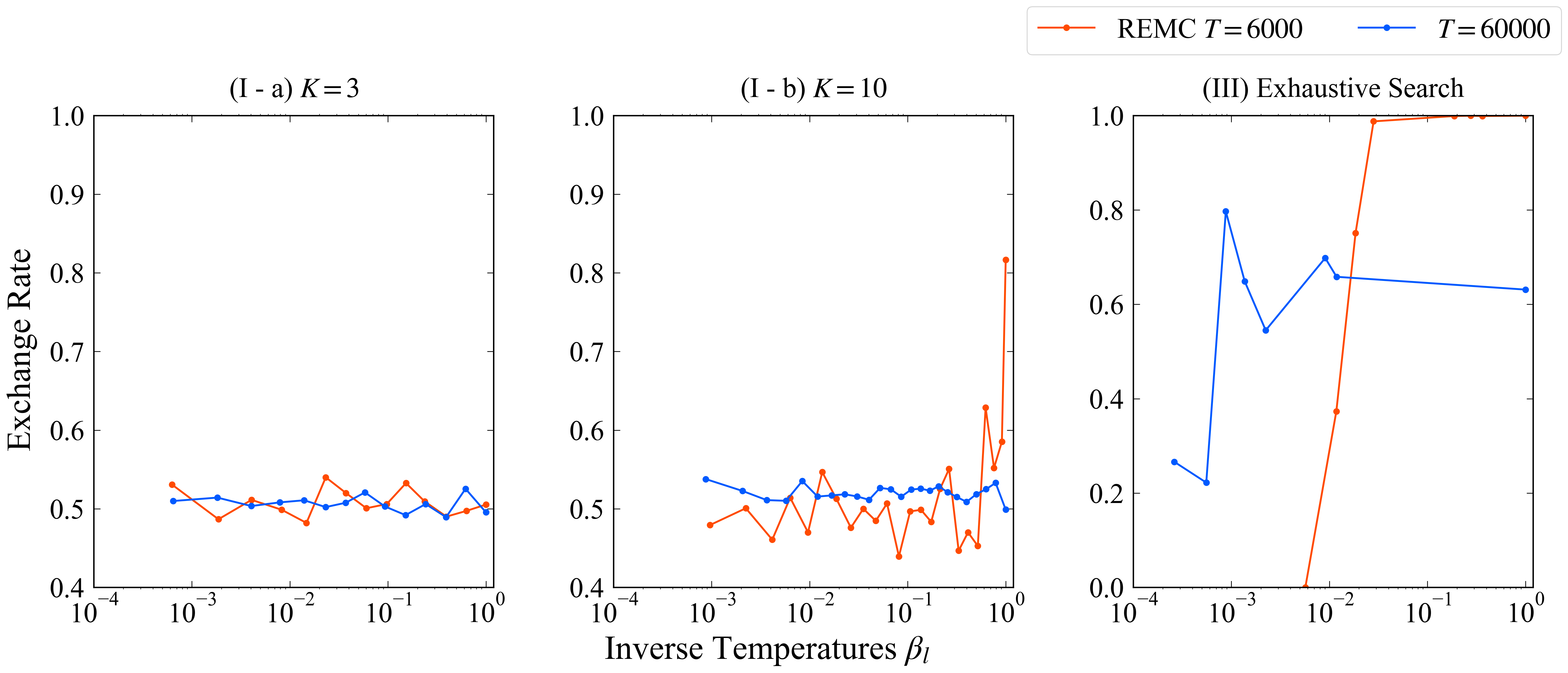}
  \caption{Exchange rates of the Metropolis algorithm at each inverse temperature for the multimodal distribution sampling with $J = 0.5$.
  From left to right, results are shown for correlation parameters $r = 0, 0.5, 0.9$.
  The red line represents the exchange rate for $T = 6000$, and the blue line represents the exchange rate for $T = 60000$.}
  \label{fig:exchange_multi_REMC}
\end{figure}

\begin{figure}[h]
  \centering
  \includegraphics[width=16cm]{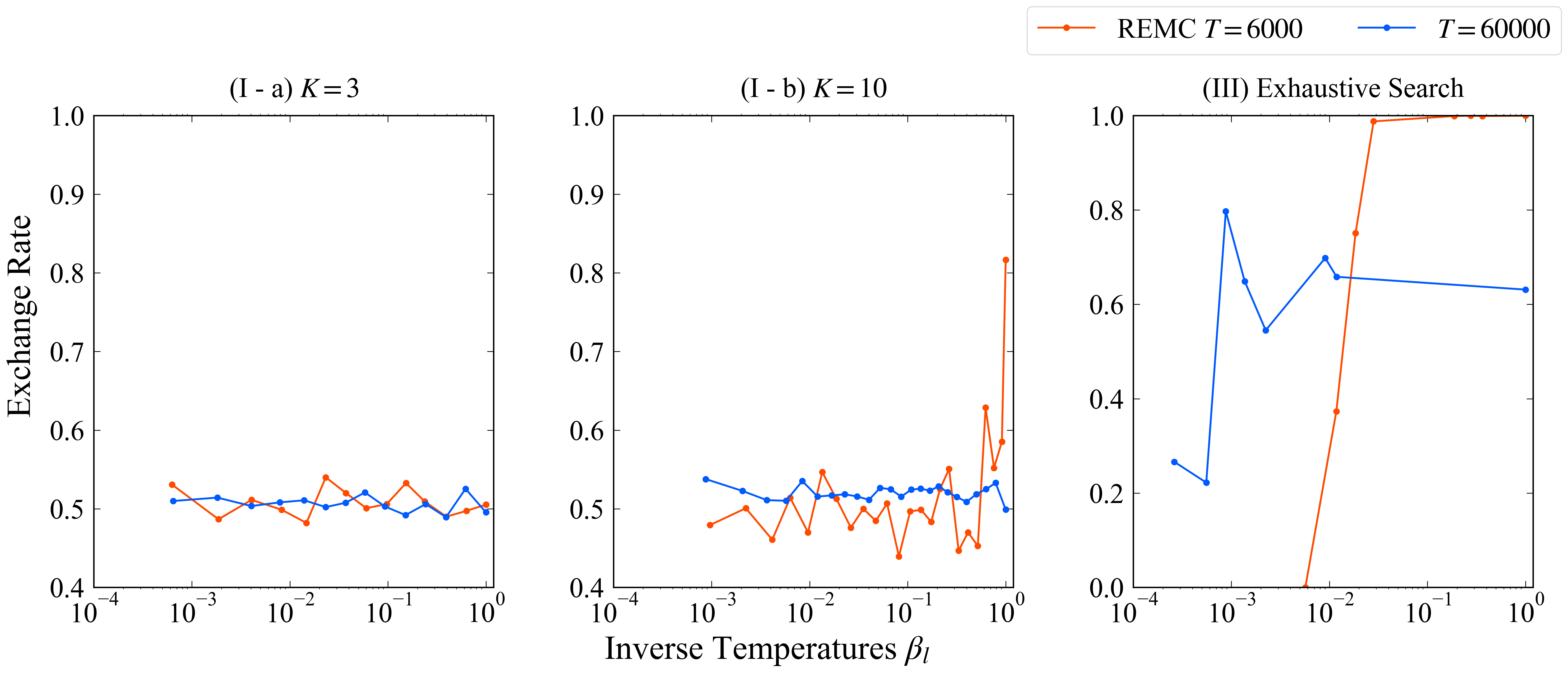}
  \caption{Exchange rates of the Metropolis algorithm at each inverse temperature for Bayesian spectral deconvolution with $J = 0.5$.
  The left and right panels correspond to the results for spectral deconvolution with $K = 3$ and $K = 10$, respectively.
  The red line represents the exchange rate for $T = 6000$, and the blue line represents the exchange rate for $T = 60000$.}
  \label{fig:exchange_bayes_REMC}
\end{figure}

We also tuned the inverse temperatures $\{\beta_l\}_{l=1}^L$ using the algorithm proposed by \cite{syed2022non}.
Figure \ref{fig:exchange_multi_REMC} and \ref{fig:exchange_bayes_REMC} show the exchange rates of the Metropolis algorithm at each temperature.
The sample sizes were set to 6{,}000 and 60{,}000, with a burn-in period corresponding to 50\% of the samples.
The results indicate that the exchange rate remained around 0.5 across all temperatures when the sample size was 60{,}000.

\newpage
\section{Theoretical Proof for Sequential Exchange Monte Carlo} 
\label{sec:appendix_proof}

In this section, we prove the following theorem:

\begin{theorem}
Let $\{\theta_l^i\}_{i=1}^N$ be the samples at the $l$-th intermediate distribution by SEMC.  
Then empirical measure $p_l^{N}$ converges in probability to the target distribution $p_{\beta_l}$ on partition $\Delta^{k}$.  
That is, for every $0 < \delta < 1$ and every $\epsilon > 0$, $K\in\mathbb{N}$ exists such that, for all $k \ge K$, $N_0\in\mathbb{N}$ exists such that, for all $N\ge N_0$,
\begin{align}
  \Pr\!\left( \|p_l^{N}-p_{\beta_l}\|_{\Delta^{k}} < \epsilon \right) > \delta .
\end{align}
\end{theorem}

First, set the parameter vector $\bm{\theta}_l=(\theta_{l-1},\theta_l)$.  
Let $P_l^{1}$ be the Markov kernel that samples $\theta_{l-1}$ in $\bm{\theta}_l$ from the empirical distribution $\bar{p}_{l-1}^N$ of the previous intermediate distribution,
and let $P_l^{2}$ be the Markov kernel that swaps $\theta_{l-1}$ and $\theta_l$ with the probability
\begin{align}
  \alpha_l^{1}(\theta_{l-1},\theta_l)
  =\min\!\left\{1,\,
        \frac{p_{\beta_l}(\theta_{l-1})\,p_{\beta_{l-1}}(\theta_l)}{p_{\beta_{l-1}}(\theta_{l-1})\,p_{\beta_l}(\theta_l)}
      \right\}.
\end{align}
Next, for each component $\theta_{l,i}$ of $\theta_l$, let $q^{2,i}$ be the proposal distribution of a random walk Metropolis step with uniform step size $h_l$, and let $P_l^{3,i}$ be the corresponding Markov kernel with acceptance probability
\begin{align}
  \alpha_l^{2}(\theta,\theta')
  =\min\!\left\{1,\,
        \frac{p_{\beta_l}(\theta')}{p_{\beta_l}(\theta)}
      \right\}.
\end{align}
Define the composite kernel
\begin{align}
  P_l^{3}=P_l^{3,1}\circ P_l^{3,2}\circ\cdots\circ P_l^{3,d},
\end{align}
and
\begin{align}
  Q_l \;=\; P_l^{1}\circ P_l^{2}\circ P_l^{3}.
\end{align}
$Q_l$ is the Markov kernel actually used to sample from the $l$-th intermediate distribution in SEMC.

For the proof, we introduce auxiliary kernels.  
Let $\widetilde{P}_l^{1}$ be the kernel that samples $\theta_{l-1}$ directly from the proposal $p_{\beta_{l-1}}$ (i.e., acceptance probability 1), and we put
\begin{align}
  \widetilde{Q}_l \;=\; \widetilde{P}_l^{1}\circ P_l^{2}\circ P_l^{3}.
\end{align}
Let $\widetilde{Q}_l^{\mathrm{hist}}$ and $Q_l^{\mathrm{hist}}$ be the histogram discretizations of $\widetilde{Q}_l$ and $Q_l$, respectively.  
More precisely, for any cells $A_{l-1},A_l\subset\Delta$ and all $\theta_{l-1}\in A_{l-1}$, $\theta_l\in A_l$, define the acceptance probability
\begin{align}
  \alpha_{l}^{1,\mathrm{hist}}(\theta_{l-1},\theta_l)
  =\min\!\left\{1,\,
        \frac{p_{\beta_l}(A_{l-1})\,p_{\beta_{l-1}}(A_l)}
             {p_{\beta_{l-1}}(A_{l-1})\,p_{\beta_l}(A_l)}
      \right\},
\end{align}
and denote the resulting swap kernel by $P_l^{2,\mathrm{hist}}$.  
Likewise, with the uniform random walk proposal
\begin{align}
    q^{2,i,\mathrm{hist}}(\theta, d\theta') \propto
    \begin{cases}
        1 & \text{if } \widetilde{\theta}^i_l - h_l^i \leq \widetilde{\theta}'^i \leq \widetilde{\theta}^i_l + h_l^i \ \ (\forall \widetilde{\theta'} \in A \ni \theta', \forall\widetilde{\theta} \in A \ni \theta) \\
        0 & \text{otherwise},
    \end{cases}
\end{align}
define the acceptance probability
\begin{align}
  \alpha_{l}^{2,\mathrm{hist}}(\theta,\theta')
  =\min\!\left\{1,\,
        \frac{p_{\beta_l}(A')}{p_{\beta_l}(A)}
      \right\},
\end{align}
 and the resulting Metropolis kernel $P_l^{3,\mathrm{hist}}$, and set
\begin{align}
  Q_l^{\mathrm{hist}}
  =P_l^{1}\circ P_l^{2,\mathrm{hist}}\circ P_l^{3,\mathrm{hist}},\qquad
  \widetilde{Q}_l^{\mathrm{hist}}
  =\widetilde{P}_l^{1}\circ P_l^{2,\mathrm{hist}}\circ P_l^{3,\mathrm{hist}}.
\end{align}
Both $\widetilde{Q}_l^{\mathrm{hist}}$ and $Q_l^{\mathrm{hist}}$ can be seen as Markov kernels that act on the discrete space $\Delta\times\Delta$.

\begin{lemma}[Uniform ergodicity of SEMC kernels]\label{lem:ue}
The kernels $Q_l$ and $\widetilde{Q}_l^{\mathrm{hist}}$ are uniformly ergodic.
\end{lemma}

\begin{proof}
Let $M=\sup_{\theta} p_{\beta_l}(\theta)/p_{\beta_{l-1}}(\theta)$ and $\bm{A}_l=A_{l-1}\times A_l$. ($A_{l-1},A_l$ are subsets of the parameter space.)  
Then, for all $\bm{\theta}_l$, the following holds:
\begin{align}
  & (Q_l\circ Q_l)(\bm{\theta}_l,\bm{A}_l) \\
  &\ge
    \int
      \bar{p}_{l-1}^N(d\theta_{l-1}')\,
      \alpha_l^{1}(\theta_{l-1}',\theta_l)\,
      P_l^{3}(\theta_{l-1}',d\theta_l')\,
      \bar{p}_{l-1}^N(d\theta_{l-1}'')\,
      \alpha_l^{1}(\theta_{l-1}'',\theta_l'')\,
      P_l^{3}(\theta_{l-1}'',d\theta_l'') \\
  &\ge
    \int
      \frac{1}{M}\,
      \frac{p_{\beta_l}(\theta_l')}{p_{\beta_{l-1}}(\theta_{l-1}')}\,
      \bar{p}_{l-1}^N(d\theta_{l-1}')\,
      P_l^{3}(\theta_{l-1}',d\theta_l')\,
      \bar{p}_{l-1}^N(d\theta_{l-1}'')\,
      \alpha_l^{1}(\theta_{l-1}'',\theta_l'')\,
      P_l^{3}(\theta_{l-1}'',d\theta_l'').
\end{align}
Define a probability measure
\begin{align}
  &\nu_l(\bm{A}_l) \\
  &\propto 
    \int
      \frac{1}{M}\,
      \frac{p_{\beta_l}(\theta_l')}{p_{\beta_{l-1}}(\theta_{l-1}')}\,
      \bar{p}_{l-1}^N(d\theta_{l-1}')\,
      P_l^{3}(\theta_{l-1}',d\theta_l')\,
      \bar{p}_{l-1}^N(d\theta_{l-1}'')\,
      \alpha_l^{1}(\theta_{l-1}'',\theta_l'')\,
      P_l^{3}(\theta_{l-1}'',d\theta_l'')
\end{align}
to obtain a minorization condition and hence a uniform ergodicity of $Q_l$ \cite{rosenthal1995minorization}. 
A similar argument, with sums replacing integrals, shows a uniform ergodicity of $\widetilde{Q}_l^{\mathrm{hist}}$.
\end{proof}

\begin{lemma}[Stationary distribution of the SEMC kernels]\label{lem:stationary}
Assume that for all $0 < \lambda < 1$ and $\epsilon > 0$,
$K\in\mathbb{N}$ exists, and for each $k\ge K$, an $N_0\in\mathbb{N}$ exists such that, for all $N\ge N_0$,
\begin{align}
  \label{eq:assump}
  \Pr\left(\|\bar{p}_{l-1}^{N}-p_{\beta_{l-1}}\|_{\Delta^{k}}<\epsilon\right) > \lambda,
\end{align}
where $\bar{p}_{l-1}^{N}$ is the empirical measure of samples $\{\theta_{l-1}^i\}_{i=1}^N$ drawn from $Q_{l-1}$.
Then the empirical measures $p_{l-1}^{N}$ of samples $\{\theta_{l-1}^i\}_{i=1}^N$ and $\bar{p}_l^{N}$ of samples $\{\theta_l^i\}_{i=1}^N$ obtained from $Q_l$ converge in probability to the target distributions $p_{\beta_{l-1}}$ and $p_{\beta_l}$, respectively, which means that for every $0 < \lambda < 1$ and $\epsilon > 0$, $K\in\mathbb{N}$ exists and, for each $k\ge K$, $N_0\in\mathbb{N}$ exists such that, for all $N\ge N_0$,
\begin{align}
  \Pr\ \!\bigl(\|p_{l-1}^{N}-p_{\beta_{l-1}}\|_{\Delta^{k}}<\epsilon\bigr)>\lambda,
  \qquad
  \Pr\ \!\bigl(\|\bar{p}_l^{N}-p_{\beta_l}\|_{\Delta^{k}}<\epsilon\bigr)>\lambda.
\end{align}
\end{lemma}

\begin{proof}
Let $\lambda, \epsilon$ be arbitrary positive numbers with $0 < \lambda < 1$.
Let $\widetilde{\pi}_l=p_{\beta_{l-1}}\times p_{\beta_l}$ be the stationary distribution of $\widetilde{Q}_l$, and let $\widetilde{\pi}_l^{\mathrm{hist}}$ be that of $\widetilde{Q}_l^{\mathrm{hist}}$.
From the detailed balance condition of $\widetilde{Q}_l^{\mathrm{hist}}$, we have
\begin{align}
  \widetilde{\pi}_l^{\mathrm{hist}}(\bm{\theta})
  =\frac{1}{\mathrm{vol}(\bm{A})}\int_{\bm{A}}\widetilde{\pi}_{\beta_l}(\bm{\theta})\,d\bm{\theta},
  \qquad \bm{\theta}\in\bm{A}\subset\Delta^{k}\times\Delta^{k}.
\end{align}
Denote the Dobrushin contraction coefficient of $\widetilde{Q}_l^{\mathrm{hist}}$ by $\delta_l^{\mathrm{hist}}$.  
By the assumption on $\bar{p}_{l-1}^N$, Lemma~\ref{lem:ue} and Eq~\eqref{eq:assump}, the following holds for $k \ge K$, $n \ge N_0(k)$ with a probability of at least $\lambda' > \lambda$:
\begin{align}
  \label{eq:bound1}
  \|\pi_l^{\mathrm{hist}}-\widetilde{\pi}_l^{\mathrm{hist}}\|_{\Delta^{k}\times\Delta^{k}}
  \;\le\;
  \frac{1}{1-\delta_l^{\mathrm{hist}}}\,
  \|\widetilde{Q}_l^{\mathrm{hist}}(\bm{\theta},\cdot)-Q_l^{\mathrm{hist}}(\bm{\theta},\cdot)\|_{\Delta^{k}\times\Delta^{k}}
  \;<\;\frac{\epsilon}{4}.
\end{align}
\par
Here, we consider the difference between $Q_l$ and $Q_l^{\mathrm{hist}}$.
Since $p_{\beta_{l-1}}$ is continuous, there exists $K^{\!*}$ such that, for every $k\ge K^{\!*}$,
\begin{align}
  \bigl|\alpha_l^{1}(\theta,\theta')-\alpha_l^{1,\mathrm{hist}}(\theta,\theta')\bigr|
  < \frac{\epsilon(1-\delta_l)}{12},
  \quad
  \bigl|\alpha_l^{2}(\theta,\theta')-\alpha_l^{2,\mathrm{hist}}(\theta,\theta')\bigr|
  < \frac{\epsilon(1-\delta_l)}{12},
\end{align}
and similarly,
$\|q^{2,i}(\theta,d\theta')-q^{2,i,\mathrm{hist}}(\theta,d\theta')\|_{\mathrm{TV}}
  <\epsilon(1-\delta_l)/12$.  
Hence, let $\delta_l$ be the Dobrushin contraction coefficient of $Q_l^{\mathrm{hist}}$; then,
\begin{align}
  \label{eq:bound2}
  \|\pi_l^{\mathrm{hist}}-\pi_l\|_{\Delta^{k}\times\Delta^{k}} < \frac{1}{1-\delta_l}\| Q_l(\bm{\theta},d\bm{\theta}')-Q_l^{\mathrm{hist}}(\bm{\theta},d\bm{\theta}')\|_{\mathrm{TV}}
  <\frac{\epsilon}{4},
\end{align}
holds.
Combining Eqs.~\eqref{eq:bound1} and \eqref{eq:bound2}, we obtain
\begin{align}
  \label{eq:bound3}
  \Pr\left(\|\widetilde{\pi}_l-\pi_l\|_{\Delta^{k}\times\Delta^{k}}
  <\frac{\epsilon}{2}\right) > \lambda'.
\end{align}

Let $\pi_l^{N}$ be the empirical distribution of the $N$ samples $\{\bm{\theta}_l^{i}\}_{i=1}^{N}$ drawn from $Q_l$.  By the law of large numbers, there exists $\widetilde{N}_0(k)$ such that, for all $N\ge \widetilde{N}_0(k)$,
\begin{align}
  \label{eq:bound4}
  \Pr\ \!\bigl(\|\pi_l^{N}-\pi_l\|_{\Delta^{k}\times\Delta^{k}}<\tfrac{\epsilon}{2}\bigr)
  > \lambda/\lambda'.
\end{align}
Taking $k\ge\max\{K,K^{\!*}\}$ and $N\ge\max\{N_0(k),\widetilde{N}_0(k)\}$, Eqs.~\eqref{eq:bound3} and \eqref{eq:bound4} yield
\begin{align}
  \Pr\ \!\bigl(\|\pi_l^{N}-\widetilde{\pi}_l\|_{\Delta^{k}\times\Delta^{k}}<\epsilon\bigr)>\lambda,
\end{align}
and the lemma follows upon marginalizing over $\theta_{l-1}$ and $\theta_l$.
\end{proof}

\begin{proof}[Proof of Theorem~1]
Let $\{(\theta_{l-1}^i,\theta_l^i)\}_{i=1}^{N}$ be the $N$ samples generated by $Q_l$, and let $p_{l-1}^{N}$ and $\bar{p}_l^{N}$ be the corresponding empirical distributions of $\theta_{l-1}$ and $\theta_l$, respectively.  
Lemma~\ref{lem:stationary} applies with $l=2$ because $P_{l=2}^{1}$ samples directly from $p_{\beta_1}$, yielding
\begin{align}
  \Pr\ \!\bigl(\|p_{1}^{N}-p_{\beta_1}\|_{\Delta^{k}}<\epsilon\bigr)>\lambda,\qquad
  \Pr\ \!\bigl(\|\bar{p}_2^{N}-p_{\beta_2}\|_{\Delta^{k}}<\epsilon\bigr)>\lambda
\end{align}
for sufficiently large $k$ and $N$.  
For $l>2$, repeated application of Lemma~\ref{lem:stationary} to successive levels establishes exactly the same bounds.  
Hence the empirical measure $p_l^{N}$ converges in probability to $p_{\beta_l}$, completing the proof.
\end{proof}
\newpage
\section{Detailed Problem Setting}
\label{sec:appendix_problem_setting}

\subsection{Free Energy of Multimodal Distribution Sampling}

This section describes the calculation of Bayesian free energy for the multimodal distribution sampling.
Since $\theta_1$ is independent of $\theta_2, \ldots, \theta_d$, the Bayesian free energy can be expressed as
$\int \exp(-E(\theta)) \textup{d}\theta = \int \exp(-\widetilde{E}(\theta_{2:d})) \textup{d}\theta_2 \cdots \textup{d}\theta_d \cdot \int \exp(-E_1(\theta_1)) \textup{d}\theta_1$,
where $E(\theta) = E_1(\theta_1) + \widetilde{E}(\theta_{2:d})$.
$\int \exp(-E_1(\theta_1)) \textup{d}\theta_1$ can be calculated as follows:
\begin{align}
    \int \exp(-E_1(\theta_1)) \textup{d}\theta_1 &= \int_0^{\frac{1}{2}} \exp(-E_1(\theta_1)) \textup{d}\theta_1 + \int_{\frac{1}{2}}^1 \exp(-E(\theta)) \textup{d}\theta_1 \\
    &= 2\int_0^{\frac{1}{4}} \exp(-30030\theta_1^2) \textup{d}\theta_1 + 2\int_{0}^{\frac{1}{4}} \exp(-30000\theta_1^2 + 15/8) \textup{d}\theta_1 \\
    &= \frac{\sqrt{\pi}}{\sqrt{30030}}\mathrm{erf}\left(\frac{\sqrt{30030}}{4}\right)
+ \exp\left(\frac{15}{8}\right)\cdot \frac{\sqrt{\pi}}{\sqrt{30000}}\,
\mathrm{erf}\!\left(\frac{\sqrt{30000}}{4}\right) 
\end{align}
where $erf(x)$ is the error function. $\int \exp(-\widetilde{E}(\theta_{2:d})) \textup{d}\theta_2, \cdots, \textup{d}\theta_d$ can be calculated as follows:
\begin{align}
    \int \exp(-\widetilde{E}(\theta_{2:d})) \textup{d}\theta_2 \cdots \textup{d}\theta_d &= \int \exp\left( - \theta_{2:d}^{\top} \bm{R} \theta_{2:d} \right) \textup{d}\theta_2 \cdots \textup{d}\theta_d \\
    &= \frac{\pi^{\frac{d-1}{2}}}{\sqrt{\det \bm R}} \\
\end{align}
where $\bm{R}$ is the correlation matrix, and $\theta_{2:d} = (\theta_2, \ldots, \theta_d)^{\top}$.
For $r = 0, 0.5, 0.9$, Bayesian free energy $F' = -\log\int \exp(-E(\theta)) \textup{d}\theta$ is calculated as $65.24, 60.18, 46.20$ respectively.

\subsection{Problem Setting for Spectral Deconvolution}
This section describes the detailed problem setting for spectral deconvolution.
For both $K=3$ and $K=10$, the number of data points was $N = 301$, and the set of inputs $X$ was $\{0.01(i-1)\}_{i=1}^N$. 
For $K = 3$, the true values of parameters and prior distributions were as follows:
\begin{align}
  \begin{pmatrix}
      a_1^*\\
      a_2^*\\
      a_3^*\\
  \end{pmatrix} 
  = 
  \begin{pmatrix}
      0.587\\
      1.522\\
      1.183\\
  \end{pmatrix}  ,\ 
  \begin{pmatrix}
      \mu_1^*\\
      \mu_2^*\\
      \mu_3^*\\
  \end{pmatrix}
  = 
  \begin{pmatrix}
      1.210\\
      1.455\\
      1.703\\
  \end{pmatrix} ,\ 
  \begin{pmatrix}
     b_1^*\\
     b_2^*\\
     b_3^*\\
  \end{pmatrix}
  = 
  \begin{pmatrix}
      95.689\\
      146.837\\
      164.469\\
  \end{pmatrix},\ 
  \sigma^2 = 0.01.
\end{align}
Let $\eta_a = 5.0, \lambda_a = 5.0, \nu_0 = 1.5, \xi_0 = 5.0, \eta_\sigma = 5.0, \lambda_\sigma = 0.04$. The prior distribution was defined as follows:
\begin{align}
  \varphi(a_k) &= \textup{Gamma} \left(a_k;\eta_a,\lambda_a\right), \\
  &= \frac{1}{\Gamma(\eta_a)}(\lambda_a)^{\eta_a}(a_k)^{\eta_a-1}\exp\left(-\lambda_a a_k\right) \\
  \varphi(\mu_k) &= N(\nu_0, (\xi_0)^{-1}), \\
  &= \sqrt{\frac{\xi_0}{2\pi}}\exp\left(-\frac{\xi_0}{2}(\mu_k-\nu_0)^2\right)\\
  \varphi(\sigma_k) &= \textup{Gamma} \left(\frac{1}{\sigma_k^2};\eta_{\sigma},\lambda_{\sigma}\right).
\end{align} \par

For $K = 10$, the true values of parameters and prior distributions were as follows:
\begin{align}
  \begin{pmatrix}
      a_1^*\\
      a_2^*\\
      a_3^*\\
      a_4^*\\
      a_5^*\\
      a_6^*\\
      a_7^*\\
      a_8^*\\
      a_9^*\\
      a_{10}^*\\
  \end{pmatrix} 
  = 
  \begin{pmatrix}
      0.587\\
      1.522\\
      1.183\\
      0.530\\
      0.300\\
      0.600\\
      0.864\\
      0.574\\
      0.388\\
      0.700\\
  \end{pmatrix}  ,\ 
  \begin{pmatrix}
      \mu_1^*\\
      \mu_2^*\\
      \mu_3^*\\
      \mu_4^*\\
      \mu_5^*\\
      \mu_6^*\\
      \mu_7^*\\
      \mu_8^*\\
      \mu_9^*\\
      \mu_{10}^*\\
  \end{pmatrix}
  = 
  \begin{pmatrix}
      0.750\\
      0.840\\
      0.950\\
      1.120\\
      1.300\\
      1.450\\
      1.710\\
      1.820\\
      1.950\\
      2.100\\
  \end{pmatrix} ,\ 
  \begin{pmatrix}
     b_1^*\\
     b_2^*\\
     b_3^*\\
      b_4^*\\
      b_5^*\\
      b_6^*\\
      b_7^*\\
      b_8^*\\
      b_9^*\\
      b_{10}^*\\
  \end{pmatrix}
  = 
  \begin{pmatrix}
      500.689\\
      1046.837\\
      764.469\\
      600.689\\
      664.469\\
      700.689\\
      550.689\\
      562.837\\
      634.469\\
      746.837\\
  \end{pmatrix},\ 
  \sigma^2 = 0.01.
\end{align}
Let $\eta_a = 5.0, \lambda_a = 5.0, \nu_0 = 1.5, \xi_0 = 1.0, \eta_\sigma = 5.0, \lambda_\sigma = 0.004$. The prior distribution was defined as follows:
\begin{align}
  \varphi(a_k) &= \textup{Gamma} \left(a_k;\eta_a,\lambda_a\right), \\
  \varphi(\mu_k) &= N(\nu_0, (\xi_0)^{-1}), \\
  \varphi(\sigma_k) &= \textup{Gamma} \left(\frac{1}{\sigma_k^2};\eta_{\sigma},\lambda_{\sigma}\right).
\end{align}

In this case, the theoretical values are unknown; therefore, we used the REMC method with a sufficiently large number of samples and distributions to obtain reference values for comparison
($T = 1{,}000{,}000$, burn-in ratio of 50\%, and $L = 50$ for $K = 3$, $T_1 = 500{,}000$, $T_2 = 500{,}000$, and $L = 50$ for $K = 10$).
The target Bayesian free energy was calculated to be $F' = 172.248, 224.364$ for $K = 3, 10$, respectively. \par

\subsection{Problem Setting for Sparse Linear Regression}
In this section, we describe the problem setting for sparse linear regression.
We consider a case with $N = 700, p = 200, s = 1, \Sigma = 0.1I$, $K = 4$, and $c_1 = 1, c_2 = 1, c_3 = 1, c_4 = 1$.
Since the theoretical values are unknown, we used the REMC method with a sufficiently large number of samples and distributions to obtain reference values for comparison with other methods
($T = 1{,}000{,}000$, burn-in ratio of 50\%, and $L = 100$).
The target Bayesian free energy was calculated as $F' = 208.956$. \par
\newpage
\section{Result for Multiple Parameter Updates}
\label{sec:appendix_result_multiple_parameter_updates}
\begin{figure}[h]
  \centering
  \includegraphics[width=12cm]{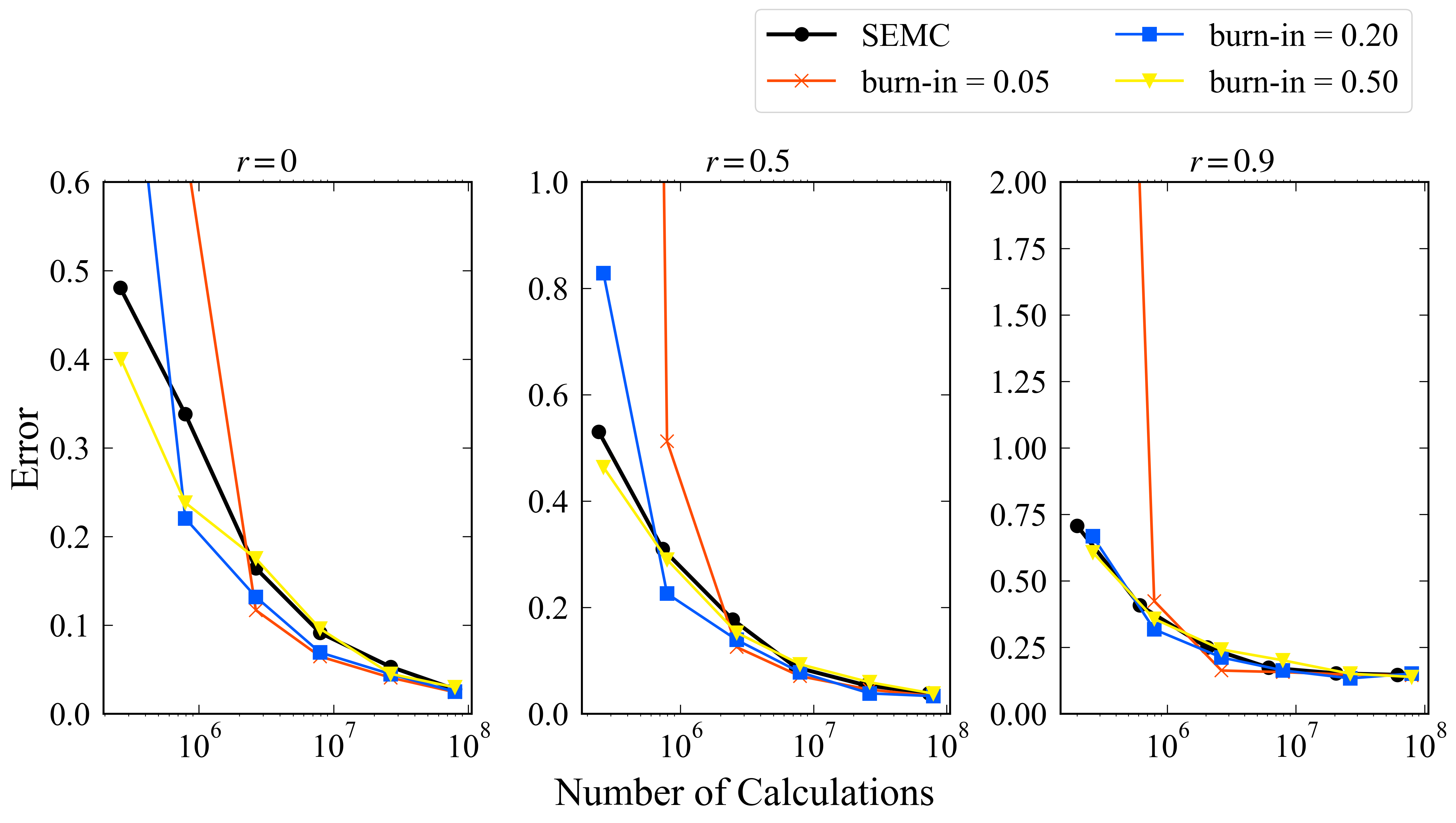}
  \caption{Comparison of SEMC and REMC in estimating the free energy under varying correlation parameters $r$.
  The mean absolute errors are shown for each method across 100 independent trials.
  From left to right, the results correspond to $r = 0, 0.5, 0.9$.
  The black line indicates the estimated free energy obtained using SEMC, while the red, blue, and yellow lines indicate the estimated free energy obtained using REMC with burn-in ratios of 5\%, 20\%, and 50\%, respectively.
  }
  \label{fig:bayes_SEMC_REMC_2}
\end{figure}

\begin{figure}[h]
  \centering
  \includegraphics[width=12cm]{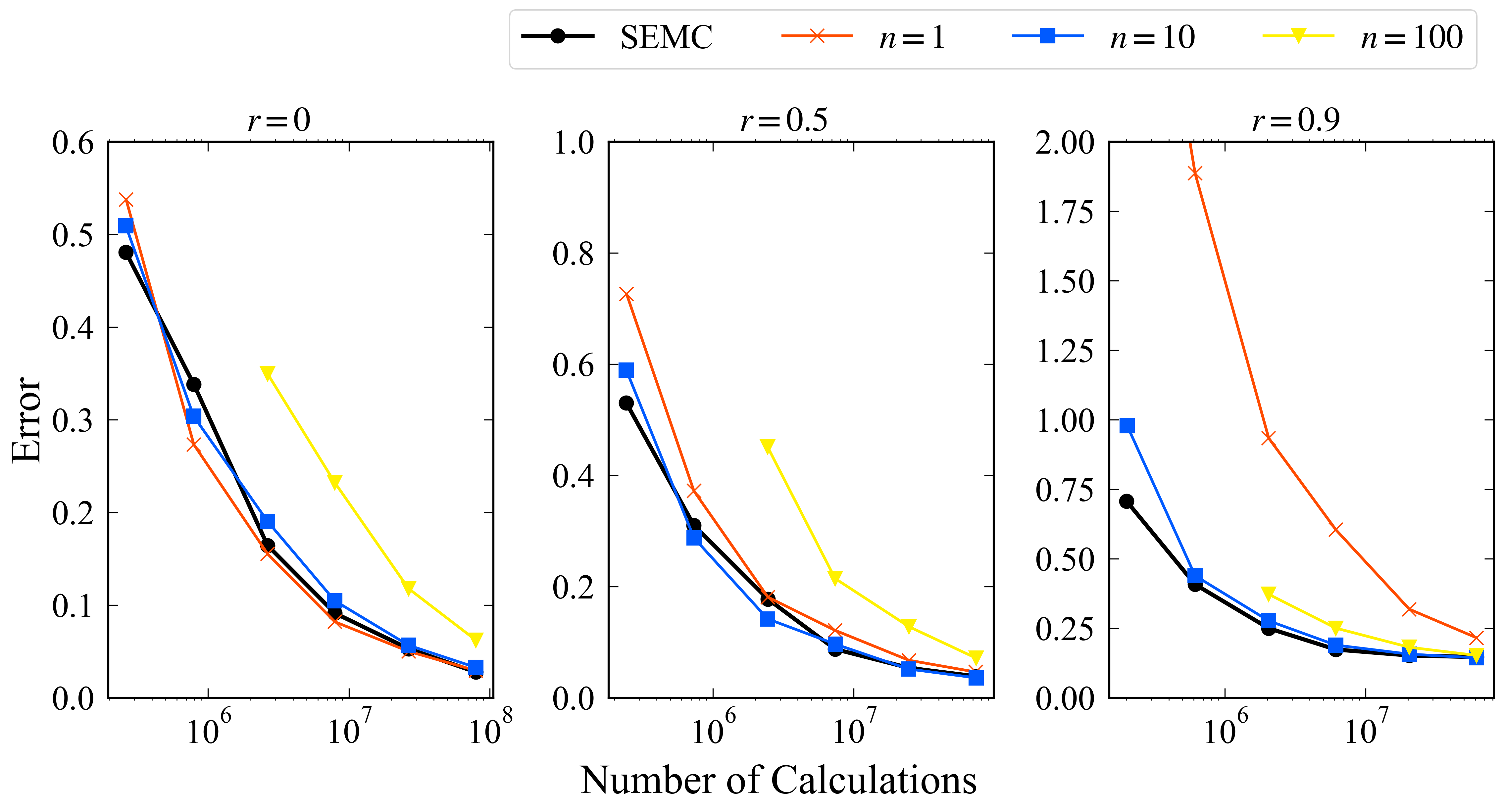}
  \caption{Comparison of SEMC and SMCS in estimating the free energy under varying sample sizes and exchange rates.
  The mean absolute errors are shown for each method across 100 independent trials.
  From left to right, the results correspond to exchange rates $J = 0.05, 0.1, 0.2, 0.5$.
  The black line indicates the estimated free energy obtained using SEMC, while the red, blue, and yellow lines indicate the estimated free energy obtained using SMCS with $n = 1, 10, 100$, respectively.
  }
  \label{fig:bayes_SEMC_SMCS_2}
\end{figure}

In this section, we present the results for multiple parameter updates using the SEMC method.
While the main text reported results for the case with a single parameter update step, here we consider the case with 10 update steps.
The corresponding results are shown in Figure~\ref{fig:bayes_SEMC_REMC_2} and Figure~\ref{fig:bayes_SEMC_SMCS_2}.
The findings indicate that the SEMC method performs comparably to the REMC and SMCS methods, even when the correlation parameter $r$ is large.

\end{document}